\documentclass[sn-mathphys-num]{sn-jnl}

\usepackage{graphicx}%
\usepackage{multirow}%
\usepackage{amsmath,amssymb,amsfonts}%
\usepackage{amsthm}%
\usepackage{mathrsfs}%
\usepackage[title]{appendix}%
\usepackage{xcolor}%
\usepackage{textcomp}%
\usepackage{manyfoot}%
\usepackage{booktabs}%
\usepackage{algorithm}%
\usepackage{algorithmicx}%
\usepackage{algpseudocode}%
\usepackage{listings}%
\usepackage{comment}
\usepackage{color,soul}
\usepackage{moresize}
\usepackage{afterpage}
\usepackage[T1]{fontenc} 
\newcommand{\pr}[1]{\left( #1 \right)}

\begin{document}

\title[Article Title]{Laboratory evaluation of a wearable instrumented headband for rotational head kinematics measurement}

\author[1]{\fnm{Anu} \sur{Tripathi}}\email{tripathia@rmu.edu}

\author[2]{\fnm{Yang} \sur{Wan}}\email{yang\_wan@brown.edu}

\author[3]{\fnm{Sushant} \sur{Malave}}\email{smalave9@gmail.com}

\author[4]{\fnm{Sheila} \sur{Turcsanyi}}\email{saturcsanyi@wisc.edu}

\author[5]{\fnm{Alice} \sur{Lux Fawzi}}\email{afawzi@wisc.edu}

\author[6]{\fnm{Alison} \sur{Brooks}}\email{brooks@ortho.wisc.edu}

\author[2]{\fnm{Haneesh} \sur{Kesari}}\email{haneesh\_kesari@brown.edu}

\author[7]{\fnm{Traci} \sur{Snedden}}\email{tsnedden@wisc.edu}

\author[4]{\fnm{Peter} \sur{Ferrazzano}}\email{ferrazzano@pediatrics.wisc.edu}

\author[5]{\fnm{Christian} \sur{Franck}}\email{cfranck@wisc.edu}

\author*[1]{\fnm{Rika} \sur{Carlsen}}\email{carlsen@rmu.edu}

\affil[1]{\orgdiv{Department of Engineering}, \orgname{Robert Morris University}, \orgaddress{\city{Moon Township}, \state{PA}, \country{USA}}}

\affil[2]{\orgdiv{School of Engineering}, \orgname{Brown University}, \orgaddress{\city{Providence}, \state{RI}, \country{USA}}}

\affil[3]{\orgname{Team Wendy}, \orgaddress{\city{Cleveland}, \state{OH}, \country{USA}}}

\affil[4]{\orgdiv{Waisman Center}, \orgname{University of Wisconsin--Madison}, \orgaddress{\city{Madison}, \state{WI}, \country{USA}}}

\affil[5]{\orgdiv{Department of Mechanical Engineering}, \orgname{University of Wisconsin--Madison}, \orgaddress{\city{Madison}, \state{WI}, \country{USA}}}

\affil[6]{\orgdiv{Department of Orthopedics and Rehabilitation}, \orgname{University of Wisconsin--Madison}, \orgaddress{\city{Madison}, \state{WI}, \country{USA}}}

\affil[7]{\orgdiv{School of Nursing}, \orgname{University of Wisconsin--Madison}, \orgaddress{\city{Madison}, \state{WI}, \country{USA}}}

\abstract{

\textbf{Purpose}\\
Mild traumatic brain injuries (mTBI) are a highly prevalent condition with heterogeneous outcomes between individuals.  A key factor governing brain tissue deformation and the risk of mTBI is the rotational kinematics of the head.
Instrumented mouthguards are a widely accepted method for measuring rotational head motions, owing to their robust sensor-skull coupling.  However, wearing mouthguards is not feasible in all situations, especially for long-term data collection. Therefore, alternative wearable devices are needed. In this study, we present an improved design and data processing scheme for an instrumented headband.\\
\textbf{Methods}\\
Our instrumented headband utilizes an array of inertial measurement units (IMUs) and a new data-processing scheme based on continuous wavelet transforms to address sources of error in the IMU measurements.  The headband performance was evaluated in the laboratory on an anthropomorphic test device, which was impacted with a soccer ball to replicate soccer heading.  \\
\textbf{Results}\\
When comparing the measured peak rotational velocities (PRV) and peak rotational accelerations (PRA) between the reference sensors and the headband for impacts to the front of the head, the correlation coefficients (\textit{r}) were 0.80 and 0.63, and the normalized root mean square error (NRMSE) values were 0.20 and 0.28, respectively.  
However, when considering all impact locations, \textit{r} dropped to 0.42 and 0.34 and NRMSE increased to 0.5 and 0.41 for PRV and PRA, respectively.  \\
\textbf{Conclusion}\\
This new instrumented headband improves upon previous headband designs in reconstructing the rotational head kinematics resulting from frontal soccer ball impacts, providing a potential alternative to instrumented mouthguards.
}

\keywords{Mild traumatic brain injury, Instrumented headband, Sensor array, Soccer headers, Continuous wavelet transform}

\maketitle

\section{Introduction}\label{Intro}

Mild traumatic brain injury (mTBI) is a highly prevalent condition, which is often termed a ‘silent epidemic’ since most injuries remain undiagnosed due to a lack of apparent symptoms \cite{buck2011mild}. These injuries can be caused by impacts to the head or sudden accelerations or decelerations of the head. An accumulation of concussive and sub-concussive impacts over time have been found to lead to neurodegenerative diseases such as chronic traumatic encephalopathy (CTE), dementia, and Parkinson's \cite{stein2015concussion, brett2022traumatic}. To understand potentially injurious head impact scenarios and safety limits for mTBI, long-term studies need to be conducted in environments where repeated impacts to the head are likely. Previous studies have found that large rotational head motions can cause excessive brain tissue deformation leading to neuronal injury \cite{Zhan2021, Pellman}. 
Therefore, there is a critical need for rotational head kinematics data collection over long time periods and in diverse situations to inform us of the potential brain injury risk. 
 
The accuracy of collected head kinematics data is dependent on the measurement system used to collect the data. Presently, instrumented custom mouthguards have been demonstrated to be the most accurate wearable sensor system in measuring head kinematics, both in the laboratory as well as on the field \cite{Karton, Buice2018, Tiernan2018, Cecchi2020, patton2021, hanlon2010validation, sandmo2019evaluation, schussler2017comparison, kieffer2020two, stitt2021laboratory, rich2019development, miller2018validation, kuo2016}. However, studies have reported low participant compliance in wearing mouthguards in sports and military applications, due to communication, breathing, and comfort issues \cite{patton2021, mcguine2020does, hawn2002enforcement, boffano2012rugby, matalon2008compliance, liew2014factors, roberts2023sports, shoreinvestigation}. 
This low compliance in wearing mouthguards, especially during long-term use over a soccer season or a series of matches and practices, can prevent the collection of critical head kinematics data needed to assess injury risk and safe exposure limits  \cite{boffano2012rugby, matalon2008compliance, liew2014factors, roberts2023sports, shoreinvestigation}. Therefore, there is a crucial need for an alternative sensor system for activities where wearing mouthguards is not feasible or difficult to enforce. 
 
Several alternative wearable sensor systems are available for head kinematics measurements, including instrumented headbands and skin patches \cite{kieffer2020two, Buice2018, patton2021, Cecchi2020, hanlon2010validation, carey2021video, schussler2017comparison, tiernan2019evaluation}. These systems have been shown to have significantly lower accuracy in head kinematics measurements compared to mouthguard sensors. Poor skull–sensor coupling resulting from soft tissue deformation, slipping, and vibrations of the sensor mount can lead to spurious signals in the head kinematics measurements \cite{patton2021}. Also, the sensor capabilities, such as the intrinsic frequency response of the sensor, maximum sensing limit, trigger threshold for recording impacts, and length of signal recording, affect the data quality \cite{huber2021laboratory,wu2016bandwidth}. The filtering algorithms used in many sensor systems also remain proprietary \cite{huber2021laboratory}, limiting advancements on that front. This study aims to address some of the limitations of current sensor systems to provide a viable alternative to instrumented mouthguards. 

\begin{figure}[h]
\centering
\includegraphics[width=0.98\textwidth]{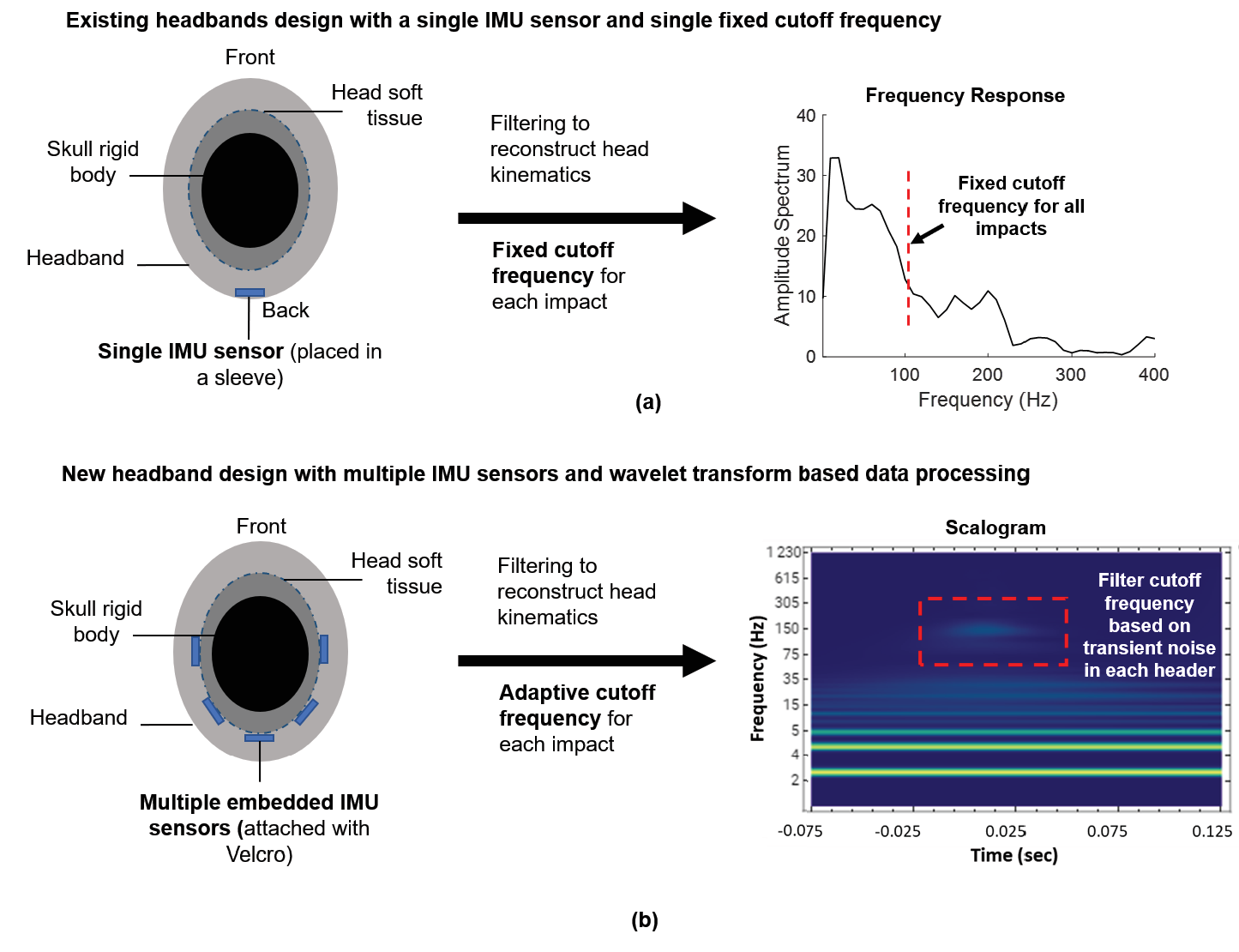}
\caption{ a) Schematic of existing headband designs, which utilize a single IMU sensor for reconstructing the head kinematics and use a single cutoff frequency when filtering the data.  b) New headband design, which utilizes an array of IMU sensors and makes use of a new continuous wavelet transform based filtering scheme where a unique cutoff frequency is defined for each head impact.}\label{fig1}
\end{figure}

In this study, we select headbands as our sensing medium and make several enhancements to their design and data processing scheme for an improved rotational head kinematics reconstruction. 
As shown in Figure \ref{fig1}a, most existing instrumented headbands have a single sensor measuring localized kinematics, which can be contaminated by soft tissue (largely skin) and headband deformations \cite{Cecchi2020,patton2021,kieffer2020two}. An array of sensors placed over a larger region of the head can provide improved head kinematics reconstruction by removing the effect of localized vibrations \cite{kuo2018head,wan2022determining}. Therefore, in this study, we have instrumented a commercially available headband with five Blue Trident sensors (Figure \ref{fig1}b). The sensors were restrained against free surface vibrations by embedding the sensors within the headband and placing them directly against the head. The sensor-headband coupling was achieved by securely attaching the sensors to the headband with Velcro. 
A new adaptive data processing method based on continuous wavelet transform was also developed to reconstruct the rotational head kinematics, where an appropriate cutoff frequency was determined for each head impact based on the signal characteristics. This differs from the filtering scheme typically used in headband studies, where a single cutoff frequency is used to filter the data.  The major differences between the new headband design and existing headbands are summarized in Figure \ref{fig1}.  

The emphasis of this study is on the rotational head kinematics reconstruction, which is the primary governing factor in brain injury \cite{takhounts2013development,zhang2006role}. We demonstrate the ability of the new headband sensor system in capturing the head kinematics of soccer headers, where rotational kinematics have raised concerns for the potential of long-term neurodegeneration \cite{Ling2017,Hales2307}.  
In this paper, we present the laboratory evaluation of the headband.  The headband evaluation tests were conducted on an Anthropomorphic Test Device (ATD), which was impacted with a soccer ball at various locations. The measured rotational velocity from the instrumented headband was post-processed using our new adaptive filter, and the resulting measurements were compared to the measurements obtained from a reference sensor embedded in the ATD to evaluate the headband performance. 

The paper is organized as follows. Section \ref{sec:method} describes the materials and methods used in the laboratory experiments and in the data processing. Section \ref{Results} provides the headband kinematics results from the experiments, and Section \ref{sec:Discussion} describes the comparison of the current headband with existing ones, as well as limitations that need to be addressed in future studies.

\section{Materials and Methods}\label{sec:method}
In this section, we describe the design of the new instrumented headband (Section \ref{sec:InstrumentedHeadband}), the experimental setup (Section \ref{sec:Experiments}), and the data processing method for the rotational head kinematics reconstruction (Section \ref{sec:DataProcessing}). 

\subsection{Instrumented Headband}
\label{sec:InstrumentedHeadband}
\begin{figure}[h]
\centering
\includegraphics[width=0.98\textwidth]{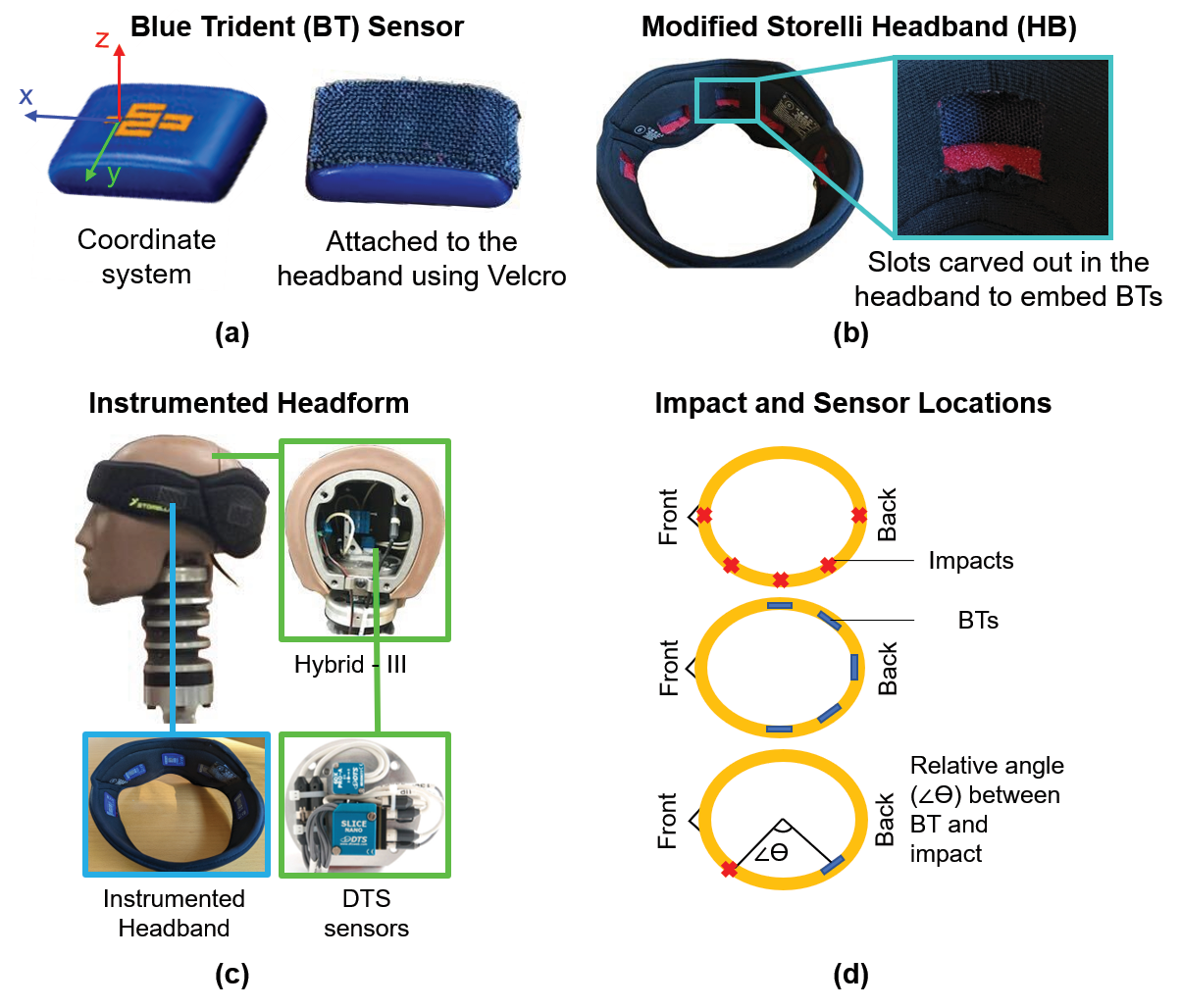}
\caption{a) Blue Trident (BT) sensors with the local coordinate system and Velcro attachment; b) Modified Storelli headband with cut-outs where each BT is attached; c) Hybrid-III headform fitted with the headband and the reference DTS sensors; and d) Top view of the headband showing the locations of the BT sensors, soccer ball impacts, and the relative angle ($\angle\theta$) between them.}\label{fig2}
\end{figure}

Our wearable sensor system consists of a headband instrumented with five inertial measurement units (IMUs), uniformly spaced from left to right around the back of the head (occipital area) (Figure \ref{fig2}).
The commercially available soccer headband, the Storelli ExoShield (Storelli Sports, Inc.), was selected for this study based on high athlete compliance in previous soccer studies \cite{mcguine2020does}. This headband was embedded with commercially available Blue Trident (BT) IMUs (Vicon Motion Systems Limited) (Figure \ref{fig2}a) to collect the kinematics data. Five BTs, each with dimensions of 42 $\times$ 27 $\times$ 11 mm, were embedded in rectangular slots of the same dimensions carved out from the inner surface of the wearable headband (Figure \ref{fig2}b-d). The embedded BT IMUs were attached securely in these slots using Velcro and positioned in direct contact with the head surface (Figure \ref{fig2}c).

BT sensors were selected for their high sampling frequencies, accuracy, high sensor range, memory storage, and continuous signal recording. Each BT IMU consists of a 3-axis gyroscope; a high-g and a low-g tri-axial accelerometer. The high-g accelerometer captures linear accelerations up to 200 g with $\pm$ 6 g accuracy (at 1600 Hz), and the low-g accelerometer captures linear accelerations up to 16 g with $\pm$ 0.05 g accuracy (at 1125 Hz). This dual accelerometer system offers high accuracy over a large range of acceleration. The gyroscope outputs angular velocities up to 2000 deg/sec ($\pm$ 5 deg/sec) at 1125 Hz. All data is continuously recorded and stored  on the onboard memory (1 GB) for up to 12 hours through the Capture.U App (Vicon Motion Systems Limited). Since the device does not utilize a trigger to record data, no data will be lost due to missed triggers (false negatives), as is the case with  other widely used head impact monitoring devices. 

\subsection{Laboratory Experiments}
\label{sec:Experiments}

\begin{figure}[h]
\centering
\includegraphics[width=0.98\textwidth]{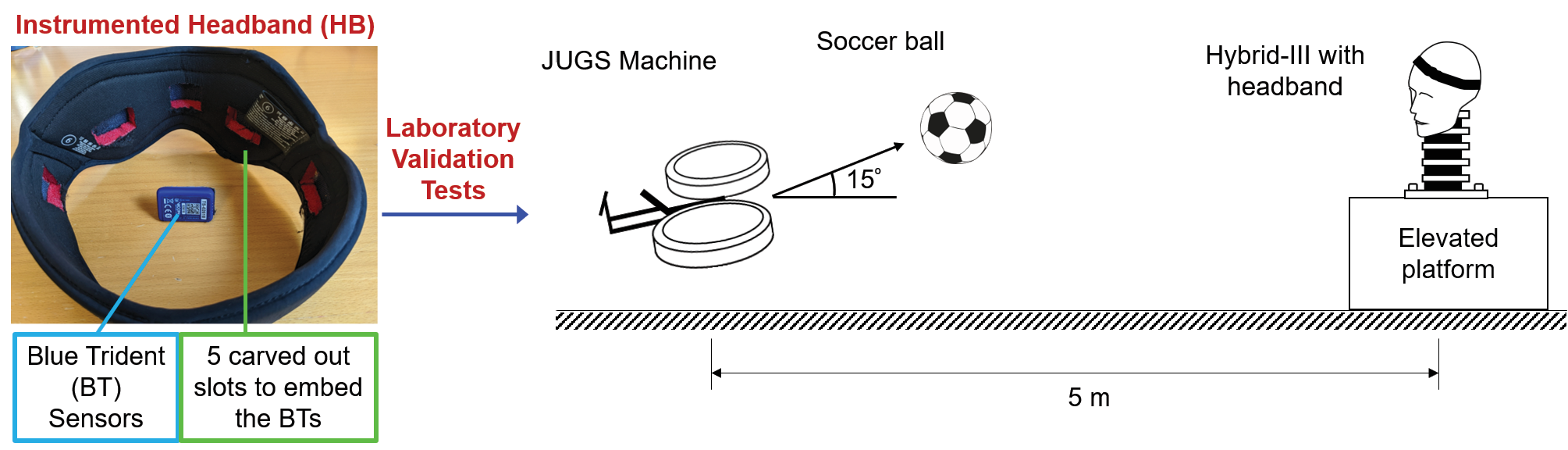}
\caption{Setup of the experimental validation tests of the new instrumented headband. An anthropomorphic test device (ATD) was impacted with soccer balls launched from a JUGS machine. The headband (HB) rotational kinematics were compared against the reference DTS sensor data for headband performance evaluation.}\label{fig3_new}
\end{figure}

The setup for the laboratory experiments is shown in Figure \ref{fig3_new}.  The headband validation experiments were conducted on an Anthropomorphic Test Device (ATD) under realistic, yet highly controlled impacts with a soccer ball to study the sources of error in the instrumented headband measurements and to reconstruct the rotational head kinematics.

A 50$^{th}$ percentile male ATD Hybrid-III head and neck form (Humanetics Innovative Solutions, Inc.) was fitted with the instrumented headband (size 6) and rigidly bolted to a flat elevated surface. The ATD was impacted with a soccer ball (Adidas size 5, $\sim$70 cm circumference), which had an internal air pressure of 84 kPa (12 psi) and was launched at a speed of 13 -- 16 m/s from a JUGS machine (JUGS Sports, Inc.) located 5 meters away from the ATD at an angle of 15° from the horizontal plane (Figure \ref{fig3_new}). The ball launch angle was selected to maintain consistency of the impact location during testing, and the ball speed was chosen to match measured ball speeds in youth soccer \cite{radja2019ball,garcia2014effects}. 

The ATD headform was fitted with a DTS 6DX PRO-A sensor (Diversified Technical Systems, Inc.) tightly attached at its center of gravity (CG) (Figure \ref{fig2}c) to provide ground truth reference head kinematics data. The reference sensor can capture 3-axes of translational accelerations up to 2000 g ($\pm$ 1-3$\%$) at 10 kHz and rotational velocities up to 8000 deg/sec ($\pm <1\%$) at 2 kHz. The DTS translational acceleration data was filtered using a low-pass Channel Filter Class (CFC) 1000 filter at 1650 Hz, and the angular velocity data was filtered using CFC 180 at 300 Hz.

Multiple locations of the head (near the front, front-side, side, back-side, and back) were impacted with the soccer ball, as shown in Figure \ref{fig2}d. The Hybrid-III head was oriented facing the JUGS machine for the frontal impacts and was rotated axially at an increment of 45° for the four successive impact locations (front-side, side, back-side, and back). The schematic in Figure \ref{fig2}d shows the location of the sensors, the impacts, and the relative angle between them ($\angle\theta$) as a measure of the sensor--impact distance. 
A total of 70 soccer ball impacts were performed (20, 20, 10, 10, and 10 impacts for front, front-side, side, back-side, and back impacts, respectively) and each impact was video recorded at 30 frames per second to identify the approximate location of the impact. 

For each impact, the head kinematics data from the headband were compared against data from the reference sensor to quantify the error in the headband data. 
This error was analyzed for impacts at different head locations and sensor--impact distances to inform the head kinematics reconstruction workflow as explained in the next section. 
Further details of the experiment and data analysis are provided in Supplementary Table S1 as per the Consensus Head Acceleration Measurement Practices (CHAMP) 2022 Reporting Guidelines on the laboratory validation of wearable sensors \cite{arbogast2022consensus}.

\subsection{Head Kinematics Reconstruction}
\label{sec:DataProcessing}

One of the challenges of measuring the rigid body kinematics of the head using an instrumented headband is that the sensors in the headband are not fixed directly to the head but are sandwiched between a material layer on the outside of the head (i.e., the ATD vinyl surface) and the soft headband foam. 
These soft materials can undergo significant deformations (Supplementary Figure S1a) as well as rigid body motion relative to the head during an impact (Supplementary Figure S1b).

\begin{figure}[h]
\centering
\includegraphics[width=0.98\textwidth]{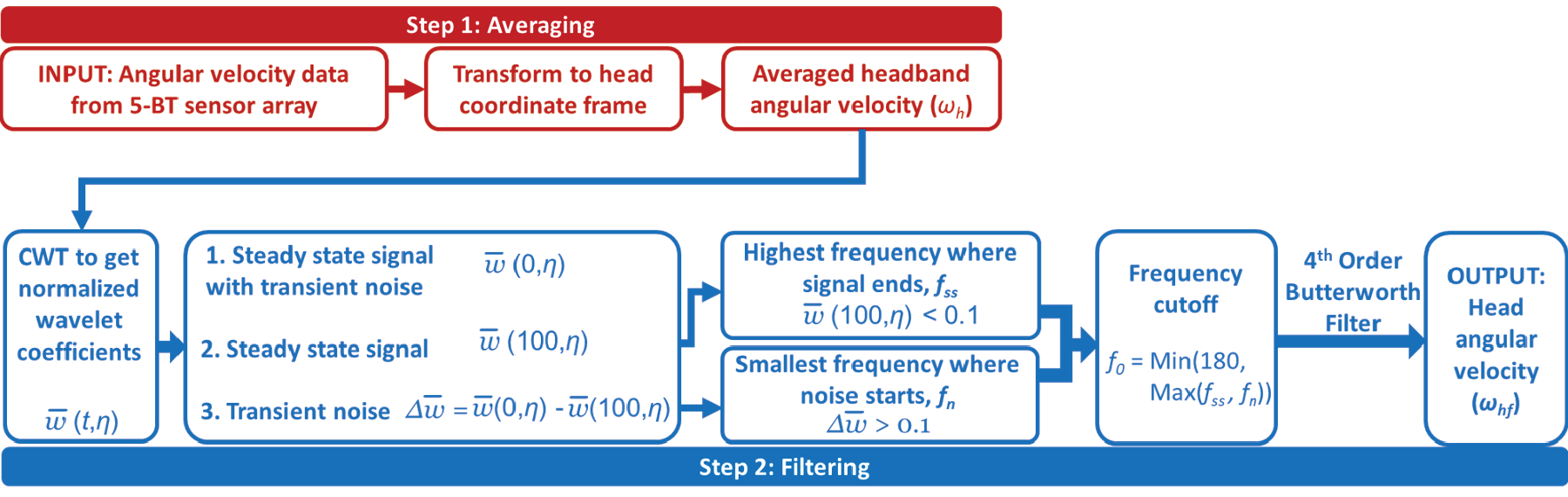}
\caption{A flowchart describing the steps involved in the averaging and the adaptive filtering based on continuous wavelet transform to reconstruct the head angular velocity. 
}\label{fig3}
\end{figure}

To remove the effect of these deformations on the measured signal and reconstruct the head angular velocity from the BT sensor data, we perform two data processing steps. First, we obtain the angular velocity of the headband by removing the contribution of the soft material deformations in the sensor measurements. This is achieved through an averaging approach using the sensor array, which is described in Section \ref{Methods:HBAngVel}. We then assume that the averaged headband angular velocity predominantly consists of the head's rigid body motion along with ``noise" from different sources, and we apply a new filtering scheme based on continuous wavelet transform to filter out this noise from the headband angular velocity and reconstruct the head angular velocity, as detailed in Section \ref{Methods:CWTFilter}. The flowchart in Figure \ref{fig3} provides an overview of the averaging and  adaptive filtering steps, which are described in the following sections.

\subsubsection{Headband Angular Velocity}\label{Methods:HBAngVel}

During a rigid body motion, the rotational kinematics of each point on a rigid body is the same. Given that the headband is not a rigid body, there are differences in the angular velocity measured at each headband sensor location (as shown in Figure \ref{fig4}a for a representative front-side impact) due to localized deformations near the sensors. Therefore, each sensor measurement consists of the headband angular velocity along with the local deformation contributions. Averaging the measurements from an array of sensors distributed over a large region can reduce the contribution of localized deformations in the signal. 

The instrumented headband contains five BT sensors distributed along the back of the head (Figure \ref{fig2}d, Section \ref{sec:InstrumentedHeadband}). 
To obtain the headband angular velocity from the sensor array, the angular velocity vectors from the five BTs were first converted to the same coordinate frame (as shown in Figure \ref{fig5}). Then, these angular velocity vectors were averaged to cancel out the deformation contributions (see Supplementary Figure S2 for an analysis of the effect of sensor-impact distance on the signal noise) and provide the headband angular velocity $\mathbf{\omega_{h}}$, which is defined as 

\begin{equation}
    \mathbf{\omega_{h}} = \frac{ \sum_{i=1}^{N} \mathbf{\omega_i} }{N} 
\end{equation}
where $\mathbf{\omega_i}$ is the angular velocity vector from each BT in the head coordinate frame, and $N$ is the number of BT sensors, which is five in this study.

Figure \ref{fig4}a shows that the averaged (unfiltered) headband angular velocity $\mathbf{\omega_{h}}$ has reduced vibrations when compared to the angular velocities measured separately from each BT sensor (BT-1 to BT-5). However, the vibrations have not been completely removed when the averaged headband angular velocity is compared to the reference DTS sensor data (Figure \ref{fig4}b).  There is still residual ``noise” in the signal.  It can be noted that the time duration of this high magnitude noise ($\sim$20--30 ms) is consistent with the contact duration of the head and soccer ball (25.16 ms, as calculated using the Hertzian--Mindlin contact model, assuming the soccer ball effective elastic modulus = 67 kPa, mass = 425 grams, speed = 13 m/s) \cite{hertz1899principles}. 
This indicates that the noise arises from the loading itself and is isolated to the headband and surrounding materials. It does not appear in the reference sensor measurements. Therefore, additional filtering must be performed, which is described in the next section.

\subsubsection{Wavelet Transform Based Filter}\label{Methods:CWTFilter}

Since small errors in the angular velocity data can get amplified when computing the angular acceleration, it is important to carefully remove this noise from the angular velocity signal while also preventing over-filtering. Here, we study the noise characteristics of the measured signals using scalograms and develop a new robust filtering scheme. 
Scalograms show a signal's frequency spectrum at any given time. A comparison of the headband and reference sensor scalograms can highlight differences in the signal characteristics, such as the frequency content, time of occurrence, and duration, informing the filtering approach. The following section briefly describes the steps to obtain a scalogram of a signal using the continuous wavelet transform (CWT). More details of the mathematics of CWT and scalograms are given in Appendix \ref{secA3}.

\paragraph{Continuous Wavelet Transform}
The CWT of the averaged headband angular velocity $\omega_h(\tau)$ is defined as
\begin{equation}
    w(\beta,\eta; \omega_{h})=\frac{1}{\sqrt{\eta}}\int_{-\infty}^{\infty}\omega_{h}(\tau)\overline{\psi}\pr{\frac{\tau-\beta}{\eta}}{\rm d}\tau,
\label{eq:wbetaeta}
\end{equation}
where 
$w(\beta,\eta;\omega_{h})$ is the wavelet coefficient at the time instance $\beta$ and frequency $\frac{f_c}{\eta \Delta\tau}$; 
$\eta$ is the scaling parameter that ranges over $\alpha 2^{{\rm oct}-1}2^{{\rm voc}/40}$, for ${\rm oct}\in \{1,2,\ldots,10\}$, ${\rm voc}\in \{1,2,\ldots,40\}$, and $\alpha=1.92$; $\Delta \tau$ is the time difference between two consecutive signal readings, and $f_c$ is the central frequency of wavelet $\psi$. 
$\overline{\psi}\pr{\frac{\tau-\beta}{\eta}}$ is the complex conjugate of the Gabor wavelet $\psi\pr{\frac{\tau-\beta}{\eta}}$  (Eq. \eqref{eq:SelectedGaborWavelet}), 
for which the central frequency is $f_c=2.39$ \cite{sadowsky1994continuous}. The CWT of the DTS angular velocity is obtained similarly.

\begin{figure}[h]
\centering
\includegraphics[width=0.9\textwidth]{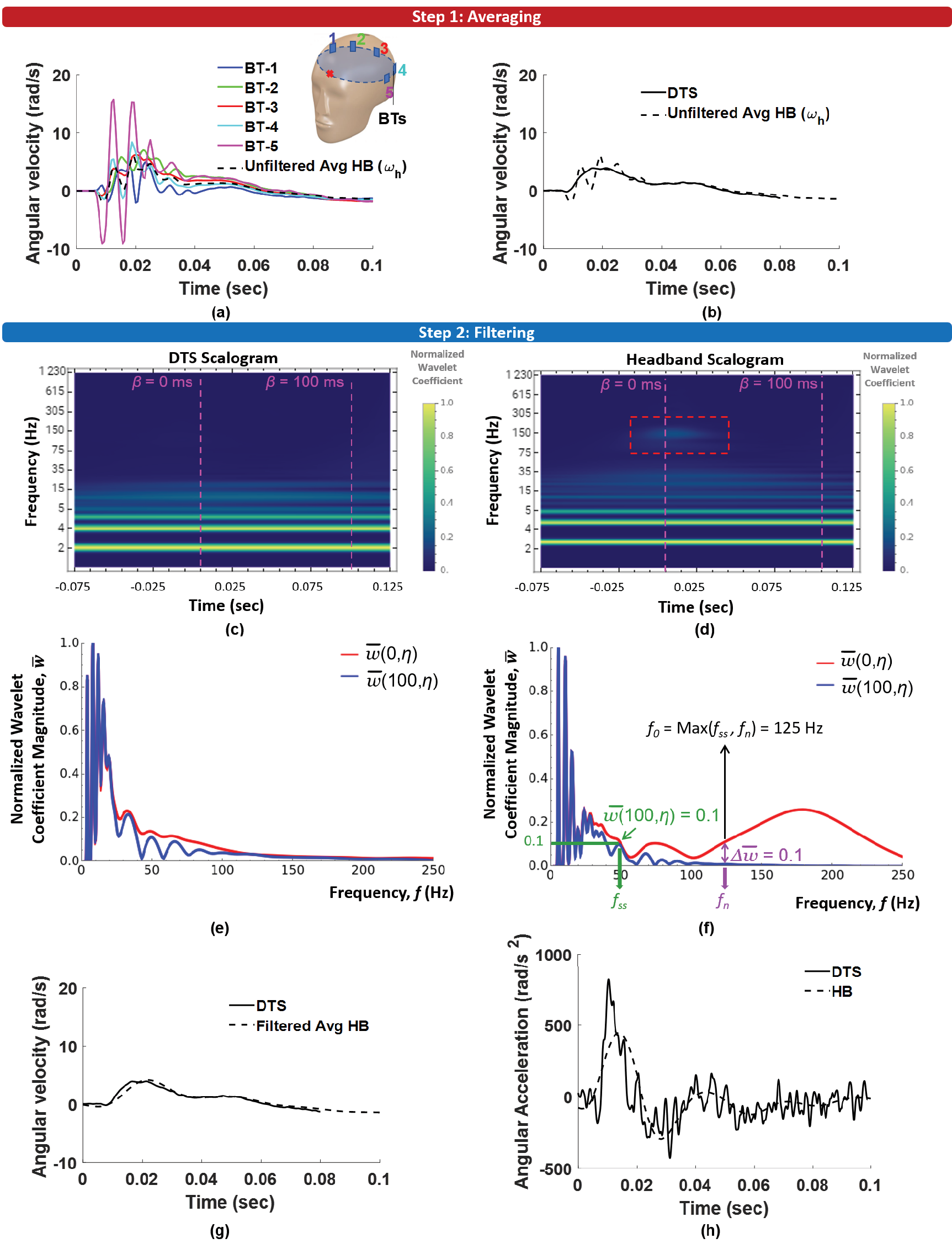}
\caption{a) Angular velocities from the five BT sensors (coronal components) for a representative front-side impact plotted against the averaged (unfiltered) headband angular velocity (\textbf{$\omega_h$}), which has reduced vibrations;
b) Unfiltered \textbf{$\omega_h$} plotted against the reference DTS measurement showing residual noise in the BT data;
c) Scalogram of the DTS angular velocity;
d) Scalogram of the averaged headband (HB) angular velocity (\textbf{$\omega_h$}), which shows transient frequency content (red box); 
e) Normalized wavelet coefficients for the DTS angular velocity at the beginning, $\beta=$0 ms, and at the end of the head motion, $\beta=$100 ms. A close overlap between the two curves indicates low transient content in the DTS signal; 
f) Normalized wavelet coefficients for \textbf{$\omega_h$} signal at $\beta=$0 and $\beta=$100 ms. The low pass filter cutoff frequency, $f_{0}$, is selected based on the characteristics of these two curves; 
g) The filtered HB angular velocity ($\omega_{hf}$) against the reference DTS data; 
h) The angular accelerations, from differentiating the DTS and filtered average HB angular velocity. 
}\label{fig4}
\end{figure}

The scalogram is a contour plot of the magnitude of the wavelet coefficients. In this study, the scalograms are plotted over a 200 ms duration, which is sufficiently long to fully capture the total duration of the impact and to also minimize edge effects \cite{lilly2017element}. 
Figures \ref{fig4}c and d show the reference DTS and the headband angular velocity scalograms for a representative header, respectively. The beginning of the head impact, which is defined as the time when the head linear acceleration exceeds 3 g, is indicated on the scalograms by the vertical line labeled $\beta$ = 0 ms. After about 100 ms, sufficiently after the contact with the soccer ball ends, the angular velocity of the head returns to zero, which is indicated on the scalogram by the vertical line labeled $\beta$ = 100 ms. 

A comparison between the two scalograms reveals transient features in the headband scalogram that are not present in the DTS scalogram. The DTS angular velocity scalogram in Figure \ref{fig4}c shows frequencies that remain throughout the entire time duration of the head motion ($\approx$ 50 - 100 ms). Unlike the DTS scalogram, the headband angular velocity scalogram has transient frequency components (i.e., present only for a short duration) in addition to the frequencies that last over the entire duration (Figure \ref{fig4}d). A red box is drawn around these transient frequencies in Figure \ref{fig4}d. We hypothesize that the frequencies that are present throughout the entire head motion comprise the head angular velocity (the signal), and the transient frequencies are the ``noise", which must be selectively removed from the signal through filtering. 

\newcommand{\norm}[1]{\lVert #1 \rVert}

\paragraph{Filtering}

Since the transient frequency content in the headband signal can differ for each head impact, we develop an adaptive method to define the appropriate cutoff frequency, $f_{0}$, for each head impact based on the CWT analysis of the headband signal. The angular velocity data is filtered at this cutoff frequency using a $4^{th}$ order Butterworth low pass filter. 

We first define the normalized wavelet coefficient magnitude at the beginning, $\bar{w}(\beta = 0\text{ ms},\eta)$, and at the end, $\bar{w}(\beta = 100\text{ ms},\eta)$, of the head motion,
\begin{equation}
    \bar{w}(0,\eta) = \frac{w(0,\eta)}{\text{Max}(w(0,\eta))} 
    \label{eq:w0}
\end{equation}
\begin{equation}
    \bar{w}(100,\eta) =\frac{w(100,\eta)}{\text{Max}(w(0,\eta))}
    \label{eq:w100}
\end{equation}

\noindent  Figure \ref{fig4}e shows the normalized wavelet coefficients for time instance $\beta$ = 0 (red curve) and $\beta$ = 100 ms (blue curve) for the DTS angular velocity data. These curves represent the change in the wavelet coefficient magnitudes as you move along the vertical lines drawn at $\beta$ = 0 ms and at $\beta$ = 100 ms in the DTS scalogram in Figure \ref{fig4}c. A close overlap between the two curves indicates low transient content in the DTS signal. 

Figure \ref{fig4}f shows the normalized wavelet coefficients for time instance $\beta$ = 0 (red curve) and $\beta$ = 100 ms (blue curve) for the averaged headband angular velocity. The difference in these two curves shows transient frequencies in the headband data. The head angular velocity (signal) frequencies are defined as the frequencies that are present at the end of the head motion, i.e., at $\beta$ = 100 ms. The highest signal frequency where $\bar{w}(100,\eta) > 0.1$ is denoted as $f_{ss}$ (as shown in green in Figure \ref{fig4}f). 

The transient noise in the headband signal is identified by taking the difference between the normalized wavelet coefficient magnitudes ($\bar{w}$) at the beginning and at the end of the head motion, 
\begin{equation}
    \Delta\bar{w} = \bar{w}(0,\eta) - \bar{w}(100,\eta)
    \label{eq:wdelta}
\end{equation}
While this difference ($\Delta\bar{w}$) is small at low frequencies for the headband data, this difference increases at higher frequencies when the transient noise becomes more prominent. We define the lowest frequency where $\Delta\bar{w} > 0.1$ as the start of the transient noise frequencies. This critical frequency is denoted as $f_n$ (as shown in purple in Figure \ref{fig4}f) and was chosen after a sensitivity analysis to ensure that the value was close to zero but large enough to not be sensitive to small numerical artifacts in the normalized wavelet coefficient. 

The low pass filter cutoff frequency, $f_{0}$, is selected to remove the noise but preserve the frequencies that are part of the angular velocity signal. There are several scenarios that may arise in the headband data.
For cases where $f_{n} > f_{ss}$, the noise frequencies do not overlap with the signal frequencies. Therefore, to remove the noise frequencies, $ f_{n}$ is selected as the cutoff frequency. This is the case for the representative header shown in Figure \ref{fig4}f. 
For cases where $f_{n} < f_{ss}$, the noise frequencies overlap with the signal frequencies. To preserve as much of the head kinematics signal frequencies, the cutoff frequency is set equal to $f_{ss}$.
For cases where the above method leads to very high cutoff frequencies ($>180$ Hz), the cutoff frequency is set equal to 180 Hz, according to SAE J211 recommendation based on the sampling frequency \cite{grenke2002digital}. This can happen for cases where the transient noise is not prominent at any frequency, i.e., $\Delta\bar{w} < 0.1$ for all measured frequencies and $f_{n}>180$ Hz. Therefore, the cutoff frequency is defined as follows:
\begin{equation}
    f_{0}= \begin{cases}
         \text{Max(}f_{ss},f_{n}), &   \text{Max(}f_{ss},f_{n}) < 180 \text{Hz}\\
         180 \text{ Hz}, &   \text{Max(}f_{ss},f_{n}) > 180 \text{Hz}.
        \end{cases}
\end{equation}

\noindent In Figure \ref{fig4}g, the filtered headband angular velocity data is plotted against the angular velocity measured from the DTS sensor for the representative header. The filtering has sufficiently removed the high frequency noise in the averaged headband angular velocity data in Figure \ref{fig4}b. The angular accelerations are calculated from the filtered angular velocity using a five-point stencil \cite{rich2019development}, and these results are shown for the representative header in Figure \ref{fig4}h. There is relatively good agreement between the headband and DTS derived angular accelerations; however, there is some noise present in both signals, which is a direct result of the numerical differentiation.

The CWT and filtering algorithms were implemented in a custom Mathematica \cite{WolframResearch} code, and further data analyses were performed in a custom MATLAB \cite{MATLAB:2020} code.
The computational cost of the algorithm is minimal, with the CPU time for wavelet-based filtering of an individual header to be $\sim$0.23 seconds when using a cloud-based Mathematica computing platform (Wolfram Research, Inc.).

\paragraph{Data Analysis}

We performed statistical analyses to compare the headband measurements of the peak rotational velocities (PRV) and peak rotational accelerations (PRA) to the reference measurements. We first normalized the reference and headband data using the maximum reference measurements for both PRV and PRA datasets and calculated the correlation coefficient ($r$). We also computed the normalized root mean square error (NRMSE) (defined in Appendix \ref{secA1}), which is less sensitive to the range of measured data. Higher values of NRMSE indicate larger error between the headband and reference sensor data. Since the concordance correlation coefficient (CCC) has been used in recent studies to quantify the agreement between sensor and reference measurements \cite{lawrence1989concordance, kieffer2020two}, we also computed the CCC values for PRV and PRA separately as defined in Appendix \ref{secA1}. Higher $r$ and CCC values indicate greater agreement between the data sets.  

We also conducted Bland-Altman analyses on both normalized PRV and normalized PRA to understand the bias between the headband and the reference sensor measurements. The peak value of the headband data for each header was subtracted from the peak value of the reference data. The average of this difference provides the mean bias, and the mean $\pm$ $1.96\times$standard deviation (SD) of the difference provides the limits of agreement, within which $95\%$ of the measurements is expected to fall. A lower mean and SD of the bias would indicate that the two methods can be used interchangeably. 

Finally, we conducted a CORA analysis to quantify agreement in the kinematic data time histories using recommended CORA parameters \cite{giordano2016development}. The angular velocity time history from 0 to 100 ms was used in the CORA analysis. Based on the CORA score, the biofidelity ratings are defined as `excellent' ($>0.86$), `good' ($0.66-0.86$), `fair' ($0.44-0.65$), `marginal' ($0.26-0.44$), and `unacceptable' ($<0.26$). 

\section{Results}\label{Results}
\label{sec:Results}

\begin{figure}[h]
\centering
\includegraphics[width=0.98\textwidth]{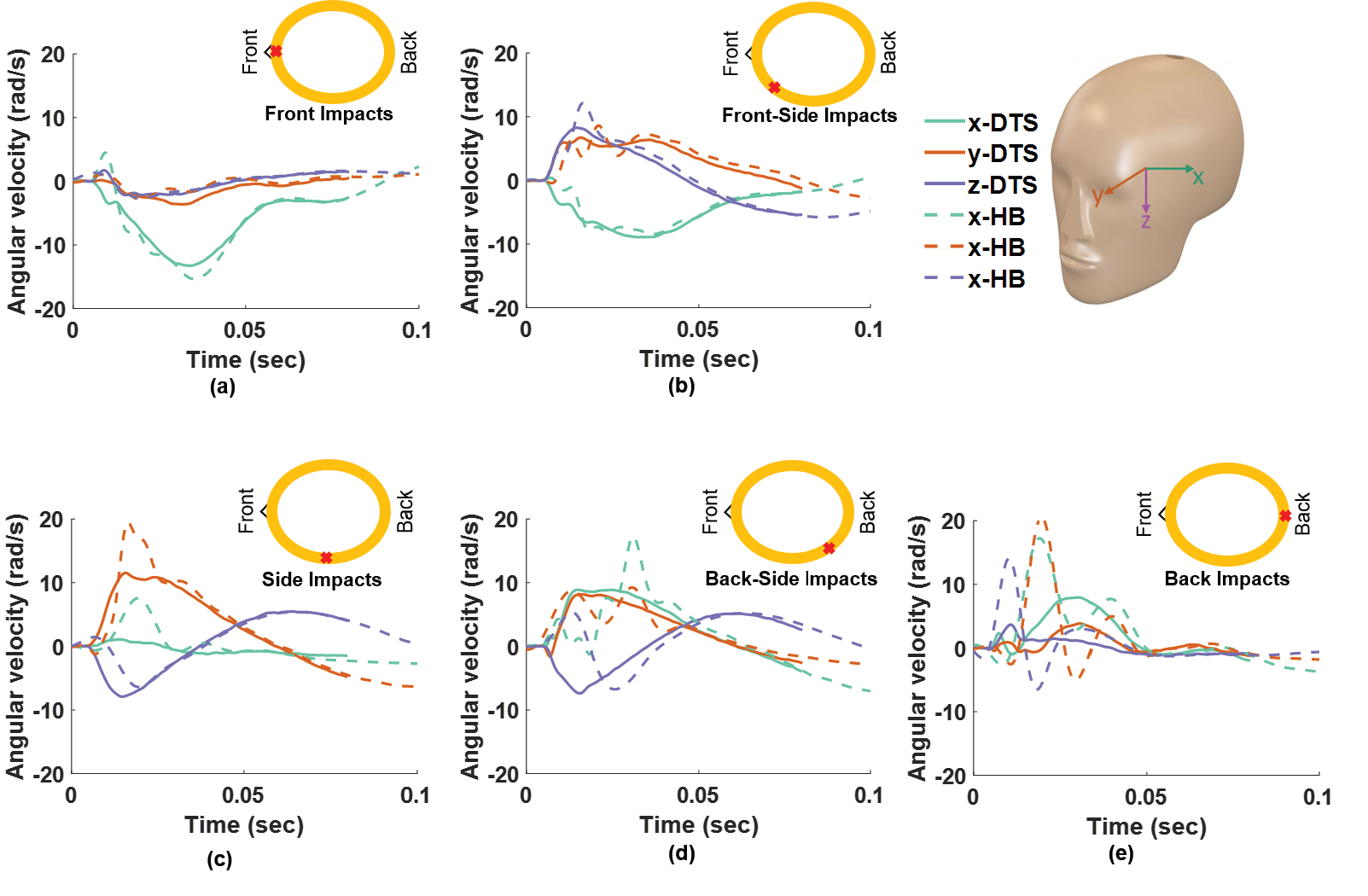}
\caption{Comparison of the DTS angular velocity time histories with the reconstructed instrumented headband angular velocity for representative headers at different head impact locations: a) front, b) front-side, c) side, d) back-side, and e) back. The impact location is indicated by the red ``x". These plots show that the head kinematics reconstruction deteriorates as the impact location moves toward the back of the head, where the sensors are located. 
}\label{fig5}
\end{figure}

A total of 70 soccer ball impacts at five different locations on the ATD head were conducted in the laboratory tests. In each impact, the soccer ball made direct contact with the headband.  The peak rotational velocities (PRV), peak rotational accelerations (PRA), and peak linear accelerations (PLA) measured by the reference DTS sensor ranged from $4.5-21$ rad/s, $700-4766$ rad/s$^2$, and $100-475$ m/s$^2$, respectively in these tests. 
The head angular velocities were reconstructed from the headband sensor data using the methods described in Section \ref{sec:DataProcessing}. 
Figures \ref{fig5}a-e compare the reference DTS angular velocity time histories with the reconstruction from the instrumented headband for representative headers at the front, front-side, side, back-side, and back locations of the head. 
These results show good agreement between the reference and filtered headband angular velocity time history curves for the front and front-side impacts. There is poorer agreement between the reference and filtered headband when the impact location is towards the back of the head where the sensors are located. 

The agreement between the reference and headband curves is quantified using the CORA scores \cite{gehre2009objective,giordano2016development}.  
The CORA scores for the 20 frontal impacts ranged between $0.70-0.92$ with average $\pm$ standard deviations of $0.82\pm0.06$, which is within the `good' ($0.65-0.86$) to `excellent' range ($>0.86$). Similarly, for the 20 front-side impacts, the CORA scores ranged from $0.72 - 0.92$ with average $\pm$ standard deviations of $0.87\pm0.06$, lying in the `good' to `excellent' range. 
On the other hand, the CORA scores for the 10 side and 10 back-side impacts ranged between $0.60 - 0.85$ ($0.70\pm0.08$) and $0.67 - 0.72$ ($0.74\pm0.04$), respectively, lying in the `good' and `fair' range. 
For the 10 back impacts, the CORA scores ranged from $0.40 - 0.67$ ($0.40 \pm 0.08$), lying in `fair' and `marginal'  ranges. 

The effect of impact location is also seen in the peak kinematic values. The correlation coefficient ($r$) between the DTS and headband measurements of the normalized PRV for impact locations moving from the front to the back of the head are 0.93, 0.51, 0.20, 0.64, and 0.01, respectively (Supplementary Figure S4a). The NRMSE for these impacts are 0.15, 0.24, 0.40, 0.33, and 1.2, respectively. The corresponding correlation coefficients for the normalized PRA are 0.58, 0.40, 0.13, -0.04, and -0.03, respectively, (Supplementary Figure S4b) and NRMSE are 0.28, 0.29, 0.40, 0.28, and 0.85, respectively. 

The Bland-Altman plots are shown in Supplementary Figure S5. The mean bias (i.e., difference) of the normalized PRV for impact locations moving from the front to the back of the head was found to be -0.07, -0.14, -0.23, -0.22, and -0.52, respectively. The mean bias values for PRA at these locations were -0.13, -0.04, -0.32, -0.41, and -0.82, respectively. These negative values indicate that the headband over-predicted the peak kinematics compared to the reference sensors, and the deviation between the DTS and the headband increases as the impact moves from the front to the back of the head. The plots show a systematic bias as the impact locations move to the back of the head. The overall mean bias values for all impact locations were -0.17 and -0.21 for PRV and PRA, respectively. The limits of agreement (lower limit, upper limit) were (-0.52, 0.18) for PRV and (-0.81, 0.39) for PRA. For impacts to the front of the head, the mean bias values were -0.09 and -0.07 for PRV and PRA, respectively. The limits of agreement were (-0.26, 0.08) for PRV and (-0.45, 0.32) for PRA for the frontal impact locations.

This sensitivity to the impact location can be explained by the sensor placement and the efficacy of the filtering method. 
When the impact location is closer to the location of the IMU sensors (i.e., the back of the head), there is poorer agreement between the angular velocities from the IMUs and the reference sensor due to more noise in the data and more overlap in the signal and noise frequencies. The noise dissipates as the distance between the impact and sensor locations increases as shown in Supplementary Figure S2. 
Since all sensors are placed towards the back of the head in our current headband design, all sensors have higher noise for impacts toward the back of the head, which is reflected in the averaged angular velocity.

The proposed filtering method analyzes the wavelet transform of the averaged angular velocity to propose an appropriate cutoff frequency for the adaptive filter. This cutoff frequency is defined as the maximum between the frequency where the transient noise begins, $f_n$, and the frequency where the signal ends, $f_{ss}$. Different factors, such as the ball inflation pressure, ball velocity, location and direction of the impact, headband material, etc., affect both the signal and the noise in the data. The proposed filtering method identifies a cutoff frequency irrespective of these conditions. 

The impact location affects the efficacy of the filtering method. As the impact moves toward the back of the head, the signal and the noise frequencies overlap (i.e., $f_n < f_{ss}$), resulting in less effective filtering (Supplementary Figure S3). For the front and the front-side impacts, the average headband angular velocity has a distinct signal-noise cutoff ($f_n > f_{ss}$), and the filtered angular velocity shows good agreement with the DTS sensor. Therefore, the instrumented headband can only accurately reconstruct the head kinematics for impacts toward the front of the head. During soccer heading, the most common head impact locations are toward the front of the head, so the headband performs well for these more common soccer heading scenarios. Changes in the soccer ball speed did not have a significant effect on the efficacy of the filtering method (Supplementary Figure S6).

\begin{figure}[h]
\centering
\includegraphics[width=0.98\textwidth]{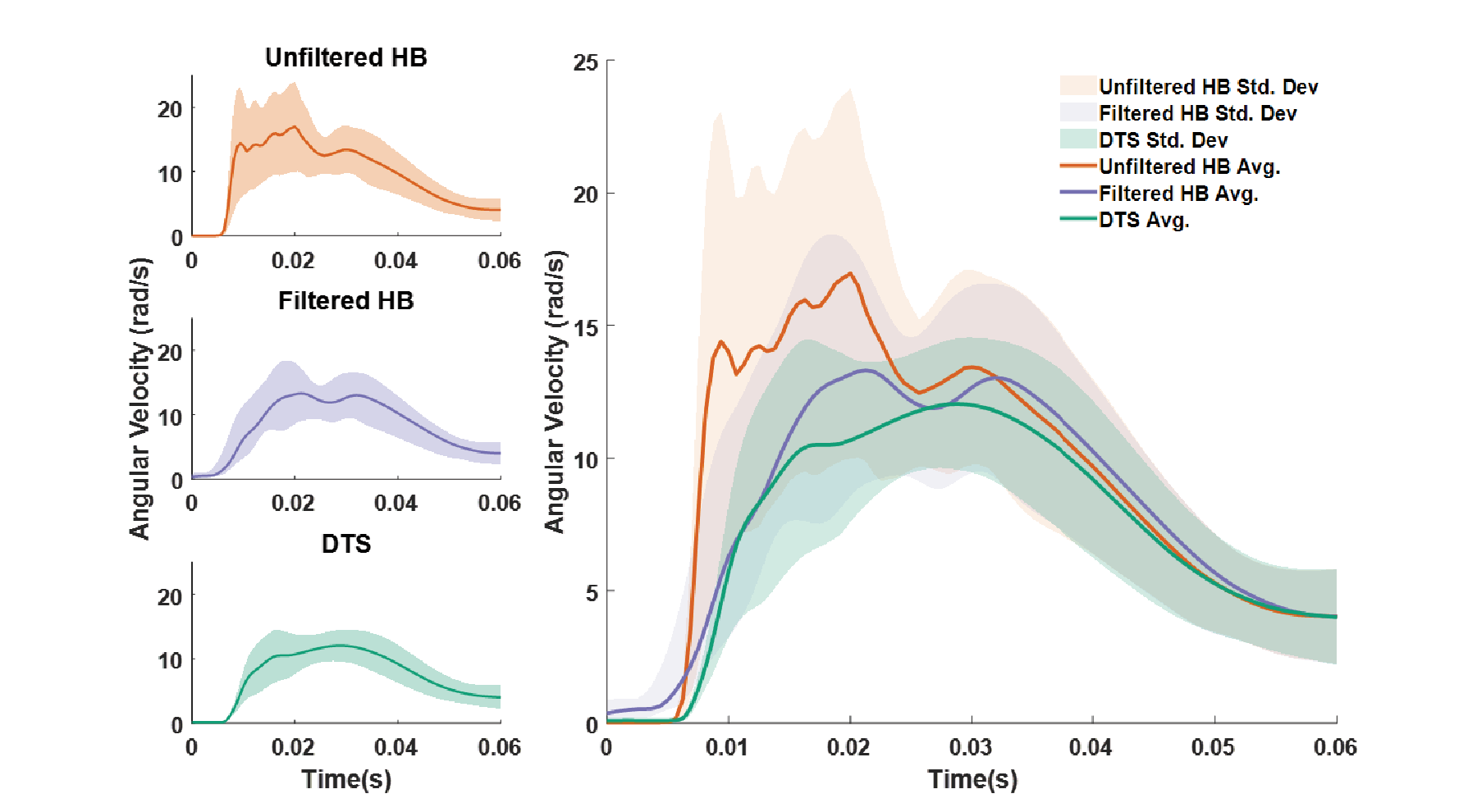}
\caption{Comparison of the average and standard deviations of the unfiltered headband (orange), filtered headband (purple), and DTS (green)  resultant angular velocities for the front and front-side impacts. There is a close overlap between the DTS and filtered headband angular velocity data for these impact locations.
}\label{fig6_new}
\end{figure}

Figure \ref{fig6_new} shows the efficacy of the filtering method on the angular velocity reconstruction. The average and standard deviations of the unfiltered headband (orange), filtered headband (purple), and DTS (green) resultant angular velocities are shown for the front and front-side impacts.  The angular velocity time history is consistently well captured for these 40 impacts using the filtered data. Further statistical analysis on the effect of filtering on the head kinematics reconstruction is provided in Supplementary Table S2, where  $r$, CCC, and NRMSE are provided for  PRV and PRA based on the unfiltered back sensor data, unfiltered averaged data, filtered averaged data using a simple Butterworth filter with a constant cutoff frequency, and filtered averaged data using the adaptive filtering method from this study. Both the averaging step and the adaptive filtering step are shown to significantly improve the statistical measures.  Furthermore, the adaptive filter is shown to outperform a simpler filter that uses a single fixed cutoff frequency.

\begin{figure}[h]
\centering
\includegraphics[width=0.98\textwidth]{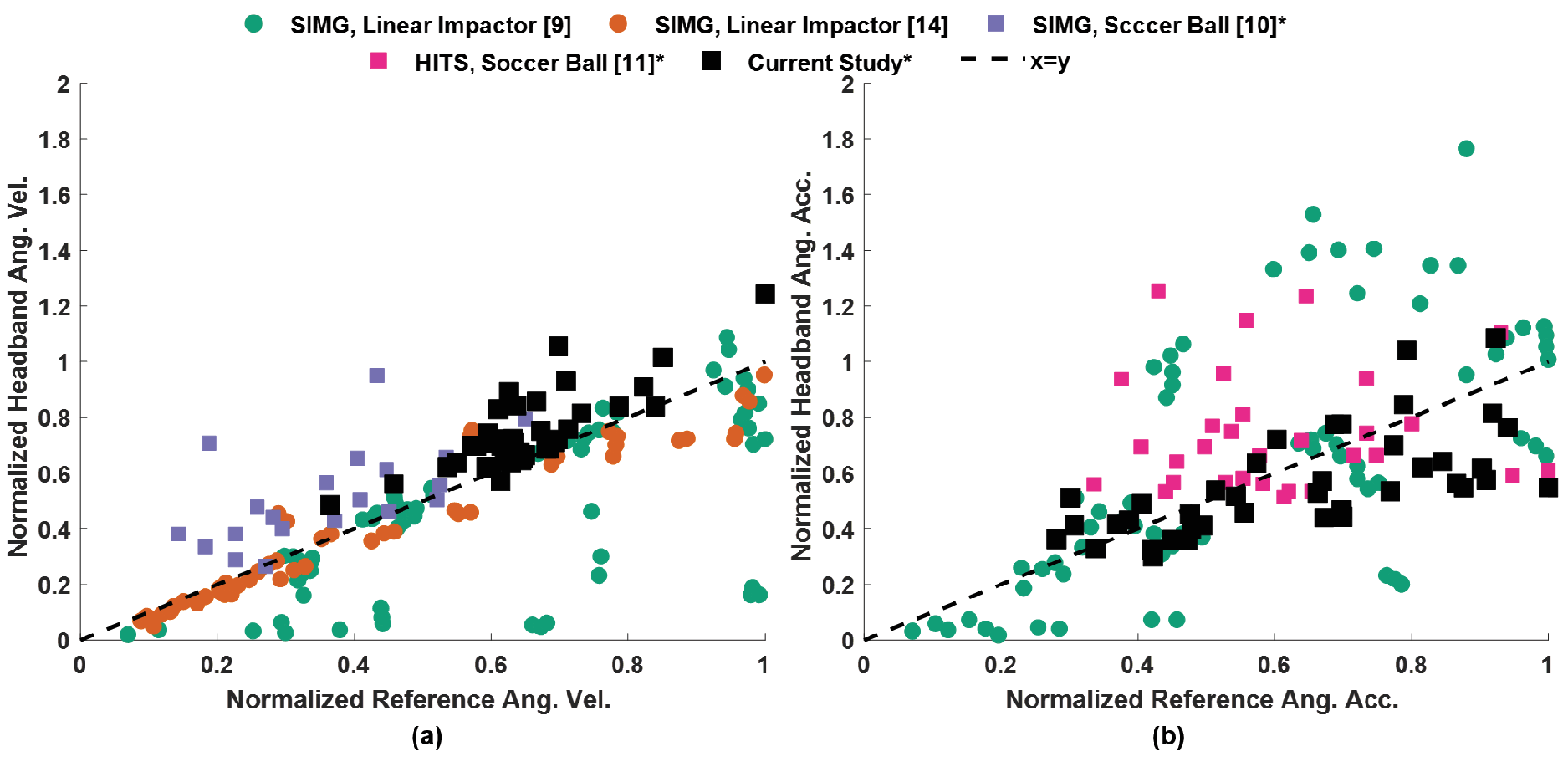}

\caption{Linear regression plots comparing the (a) peak rotational velocity (PRV) and (b) peak rotational acceleration (PRA) measured from different instrumented headbands (SIM-G, HITS, and current headband) and reference sensors for different types of impactors (soccer ball or linear impactor). For some studies \cite{Cecchi2020,kieffer2020two}, the results were not differentiated based on impact location, so the results are shown for all head impact locations.  For other studies \cite{patton2021,hanlon2010validation} and for this current study, the results are shown only for front and front-side impacts, as indicated by the asterisk (*). 
}\label{fig6}

\end{figure}

\section{Discussion}
\label{sec:Discussion}

Existing instrumented headbands have been shown to yield significantly lower accuracy in the reconstructed head kinematics compared to mouthguards, both in laboratory and field tests \cite{kieffer2020two}. However, headband data are still used in long-term exposure and computational studies primarily due to their comfort and ease of use \cite{Brooks2021, Harriss2019}. 
In this study, we used a commercially available headband and IMU sensors to construct a new instrumented headband. The signal and noise components of the headband measurements were analyzed, and a new filtering approach was developed to reconstruct the rotational head kinematics. In the following sections, we compare the results of the instrumented headband from this study against the performance of other headbands in the published literature (Section \ref{sec:Comparison}) and discuss the limitations of this study (Section \ref{sec:Limitations}).

\subsection{Comparison with existing headbands}
\label{sec:Comparison}

In this section, we compare the performance of different instrumented headbands in measuring the peak rotational velocity (PRV) and peak rotational acceleration (PRA). The existing headbands include a headband-mounted SIM-G (Triax Technologies, Norwalk, CT, USA) impact sensor and a head impact telemetry system (HITS) (Simbex, Lebanon, NH) headband.

Figure \ref{fig6} provides the linear regression plots of the PRV and PRA from these studies \cite{Cecchi2020,kieffer2020two,patton2021,hanlon2010validation}. We limit our discussion to un-helmeted impacts, which are the most similar to our current study.  For studies that provided differentiated data based on impact locations \cite{patton2021,hanlon2010validation}, only the front or front-side impact data are plotted in Figure \ref{fig6}. Otherwise, the combined data for all impact locations are plotted in Figure \ref{fig6} \cite{Cecchi2020,kieffer2020two}. The data are normalized using the maximum reference measurement from each study. The statistical correlation measures ($r$ and CCC), NRMSE, mean and standard deviation (SD) of the bias from a Bland-Altman analysis for these studies are provided in Table \ref{tab:sensor_comparison}. A higher $r$ and CCC value indicates better performance whereas a lower NRMSE, bias, and SD are desirable. Several differences in the experimental conditions make a direct comparison to the results of this study difficult, as discussed below. 

The majority of existing headband studies utilize the SIM-G headband-mounted impact sensor, whose production has been discontinued \cite{huber2021laboratory}. 
This instrumented headband utilizes a single SIM-G IMU sensor placed in a Velcro pocket towards the back of the head \cite{Cecchi2020,patton2021,kieffer2020two}. The sensor consists of a triaxial accelerometer and a triaxial gyroscope. The sensor records the linear acceleration and angular velocity at 1000 Hz after a 16 g trigger is exceeded. 

The SIM-G headband has been evaluated using a linear impactor \cite{Cecchi2020,kieffer2020two} and soccer ball impacts \cite{patton2021}. The linear impactor studies \cite{Cecchi2020,kieffer2020two} impacted an ATD headform wearing the SIM-G headband at the front, back, side, front-side, back-side, and crown of the head. All impact locations were combined in the  results of the linear impactor studies. Kieffer et al. \cite{kieffer2020two} showed very high agreement for the PRV data (high $r$ = 0.96 and CCC = 0.95; low NRMSE = 0.2, bias = 0.04, SD = 0.08) whereas Cecchi et al. \cite{Cecchi2020} consistently underestimated the PRV and showed lower agreement (lower $r$ = 0.67, CCC = 0.59; higher NRMSE = 0.45, bias = 0.15, and SD = 0.23) (Figure \ref{fig6}a). The authors attributed these errors to sensor sampling frequencies and signal processing limitations \cite{Cecchi2020}. On the other hand, Kieffer et al. \cite{kieffer2020two} report a low agreement for PRA (CCC=0.39) compared to Cecchi et al. \cite{Cecchi2020} (CCC = 0.56). In the absence of time histories, it is difficult to understand if the data-processing affected the agreement of one quantity at the expense of another, such as due to over-filtering. Some of the differences could be attributed to the use of different ATD headforms that can affect the sensor-headband coupling. It is also unclear whether the impactor made direct contact with the headband for all impact locations in these studies, which would also affect the results. Given the inconsistent results from these studies, it is difficult to conclude the performance of the SIM-G headband. Since the results were not differentiated based on the impact locations in these studies, a direct comparison to our study cannot be made.

The SIM-G headband has also been tested under soccer ball impacts, similar to our current study \cite{patton2021}. Previous studies have shown reduced performance of wearable sensors for soccer ball impacts compared to linear impactor tests \cite{sandmo2019evaluation}. Patton et al. evaluated the SIM-G headband for soccer ball impacts to the front, top (crown), side (temporal), and back (occipital) locations of a Large Omni-Directional Child (LODC) ATD headform \cite{patton2021}. For all impact locations, there are conflicting trends in the PRV data when compared to our study. This SIM-G study showed higher correlation ($r$=0.77, CCC=0.58) compared to our study ($r$=0.42, CCC = 0.25), but also showed much higher NRMSE (0.87), bias(-0.23), and SD (0.22) when compared to our study (NRMSE=0.5, bias=-0.17, SD=0.18). However, for impacts to the front of the head, our instrumented headband had improved PRV results (higher $r$=0.8, CCC=0.62; lower NRMSE=0.2, bias=-0.09, SD = 0.09) compared to the SIMG headband (lower $r$=0.58, CCC=0.38; higher NRMSE=0.89, bias=-0.16, SD = 0.14) (Figure \ref{fig6}a). The Patton study did not assess angular accelerations, and so a direct comparison to our results cannot be made for PRA.

The head impact telemetry system (HITS) (Simbex, Lebanon, NH) headband was also evaluated under soccer ball impacts and head to head impacts where the front and side of a 50th percentile male Hybrid III ATD headform was impacted \cite{hanlon2010validation}. Unlike the SIM-G headband that contains a single IMU sensor, the HITS headband was instrumented with a six single-axis linear accelerometer array placed over a small region towards the back of the head. The trigger threshold was 10 g, and the sampling frequency was 1000 Hz. The rotational accelerations were obtained using the simulated annealing optimization algorithm, which iteratively solves for the accelerations using the six linear accelerometer measurements \cite{chu2006novel}. For all impact locations, the PRA agreement is higher for the HITS headband (higher $r$=0.76, CCC=0.74; similar NRMSE=0.43, bias=0.03) when compared to our study (lower $r$=0.34, CCC=0.33; similar NRMSE=0.41, higher bias=-0.21). However, the HITS headband was not evaluated for impacts to the back of the head, which coincides with the location of the sensors and may result in poorer performance. For the impacts to the front of the head, the PRA agreement is much lower for the HITS headband (lower $r$=0.02, CCC=0.015; higher NRMSE=0.61, bias=-0.14) when compared to our study (higher $r$=0.63, CCC=0.55; lower NRMSE=0.28, bias=-0.07) (Figure 8b). The HITS study reported higher $r$ (0.95 ± 0.01) for the frontal impacts when head to head impacts were also considered, due to the expanded range of impact velocities that were tested. Our headband was not tested under head to head impacts, so a direct comparison cannot be made for these impact conditions. Also, PRV was not measured in the HITS headband study, so these results cannot be compared.

Therefore, it can be concluded that the instrumented headband presented in this study improves upon both the angular velocity and angular acceleration measurements of the head when compared to the SIM-G and HITS headbands under frontal and front-side soccer ball impacts. However, for a general impact location, the improvement in head kinematics measurement is not conclusive since some previous headband studies reported better results in either PRV or PRA \cite{patton2021,kieffer2020two,hanlon2010validation}.  There are also conflicting trends, with some studies reporting higher $r$ and CCC for the existing headbands, but also higher NRMSE, mean bias, and SD \cite{patton2021,hanlon2010validation}.  With the absence of time history curves, it is difficult to assess the effect of data-processing on the results. It should be noted that a direct comparison with some of these studies was not possible due to differences in testing conditions.

The improvements in the current headband performance can be attributed to the use of an array of sensors placed over a larger region of the head and the implementation of the new filtering scheme. Averaging the angular kinematics data from an array of sensors significantly reduces the noise in the headband angular velocity measurements. While the HITS headband used an array of single-axis accelerometers placed near the back of the head, the SIM-G headband utilized a single IMU sensor.  The new filtering approach also conserves the head kinematics signal while selectively filtering the transient frequencies. Since the noise in the signal depends on the loading conditions, such as soccer ball speed and impact location, the adaptive filtering approach enables an appropriate cutoff frequency to be uniquely identified for each impact. This filtering scheme is different from that used in prior headband studies, where a single fixed cutoff frequency is typically used to filter the kinematics signal. Although this study shows an improvement in head reconstruction with an instrumented headband, mouthguards still outperform the headband (Supplementary Table S3).

\begin{tiny}
\begin{table}[h]
    \centering
    \renewcommand{\arraystretch}{1}
    \setlength{\tabcolsep}{2.5pt}
    \small
    \begin{tabular}{llcccc|cccc}
        \toprule
        & \textbf{Impact} & \multicolumn{4}{c|}{\textbf{PRV}} & \multicolumn{4}{c}{\textbf{PRA}} \\
        \cmidrule(lr){3-6} \cmidrule(lr){7-10}
        \textbf{Sensors} & \textbf{Type} & \textbf{r} & CCC & \textbf{NRMSE} & \textbf{Bias (SD)} & \textbf{r} & \textbf{CCC} & \textbf{NRMSE} & \textbf{Bias (SD)} \\
        \midrule
         & \cite{Cecchi2020} Linear & 0.67 & 0.59 & 0.45 & 0.15 (0.23) & 0.67 & 0.56 & 0.63 & 0.04 (0.20) \\
        & Impactor  & - & - & - & - & - & - & - & - \\
        \cmidrule(lr){2-10}
        SIM-G & \cite{kieffer2020two} Linear & 0.96 & 0.95 & 0.2 & 0.04 (0.08) & - & 0.39 & - & - \\
        & Impactor & - & - & - & - & - & - & - & - \\
        \cmidrule(lr){2-10}
         & \cite{patton2021} Soccer & 0.77 & 0.58 & 0.87 & -0.23 (0.22) & - & - & - & - \\
        & Ball & 0.58* & 0.38* & 0.89* & -0.16 (0.14)* & - & - & - & - \\
        \midrule
        HITS & \cite{hanlon2010validation} Soccer & - & - & - & - & 0.76 & 0.74 & 0.43 & 0.03 (0.15)\\
        & Ball & - & - & - & - & 0.02* & 0.015* & 0.61* & -0.14 (0.26)* \\
        \midrule
        Current & Blue & 0.42 & 0.25 & 0.5 & -0.17 (0.18) & 0.34 & 0.33 & 0.41 & -0.21 (0.31) \\
        Study & Trident & 0.8* & 0.62* & 0.2* & -0.09 (0.09)* & 0.63* & 0.55* & 0.28* & -0.07 (0.20)* \\
         & IMU &  & & & & & & & \\
        \bottomrule
    \end{tabular}
    \caption{Comparison of different statistical measures for peak rotational velocity (PRV) and peak rotational accelerations (PRA). The statistical measure combines all impact locations (upper row) unless indicated by an (*), where only front or front-side impacts are considered (lower row).}
    \label{tab:sensor_comparison}
\end{table}
\end{tiny}

\subsection{Limitations}
\label{sec:Limitations}

There are several limitations of this study that should be addressed in future work. First of all, the data-processing and analyses presented in this paper are relevant to soccer ball impacts only and future laboratory studies should evaluate the headband for other types of impact, such as player-to-player, player-to-ground, multiple impact scenarios, and for a wider range of impact speeds. Other types of laboratory testing methods should also be considered.
Since the features of the measured signal can vary for different types of impacts, different data processing algorithms should  be explored to obtain accurate kinematics reconstruction.

This study evaluated the performance of the headband on an ATD in the laboratory setting; however, the head kinematics of an ATD in the laboratory setting may be quite different from that of a human player on the field. The ATD responds passively when the head is impacted with a soccer ball whereas a human player actively accelerates their head towards the ball when properly heading a soccer ball, leading to different rotational head kinematics. Also, the presence of skin, hair, and sweat can affect the sensor-skull coupling, which differs from the high-friction  ATD-sensor interface used in the laboratory tests. Therefore, field evaluation of the device on athletes is indispensable before deploying the headband for injury risk assessment in soccer. We are currently conducting a study to evaluate the headband on the field using mouthguards as the reference sensor, which will be presented in future work. In addition to evaluating the headband for soccer headers, field evaluation data should also be collected for different types of impacts. These future studies should investigate the headband performance as per the Consensus Head Acceleration Measurement Practices (CHAMP) before a large-scale deployment of this instrumented headband \cite{arbogast2022consensus,rowson2022consensus,kuo2022field}.

This study has also highlighted where further improvements can be made to the headband design. In the laboratory tests, we found that the current headband can accurately capture the rotational head kinematics for impacts to the front of the head. While soccer ball impacts to the back of the head are infrequent, there could be impact scenarios, such as player-to-player impacts, where these measurements may need to be recorded. To capture these head impacts, additional sensors could be incorporated into the headband design, covering a larger region of the head. Given that the sensors farthest away from the impact location have the highest signal-to-noise ratio, sensors should be added to the front of the head to accurately capture the rotational kinematics arising from impacts to the back of the head. Of course, it will be important to maintain the comfort and safety of the headband when incorporating additional sensors.  Incorporating more sensors will also increase the cost of the instrumented headband.  Alternately, a non-coplanar array of accelerometers can be used, which has been shown to accurately capture head kinematics measurements \cite{wan2022determining, rahaman2020accelerometer}. The method of sensor attachment to the headband can also affect the noise content (Supplementary Figure S7) and should be studied further.  

Another significant challenge of instrumented headbands is preventing sliding at the head-headband interface \cite{kieffer2020two,patton2021}. Although a tight-fitting headband will reduce the amount of sliding, it will not fully prevent sliding in high-intensity impacts to the head. The effect of headband size and tightness of fit was not evaluated in this study.  Future headbands can be designed to further reduce sliding by increasing the coefficient of friction or adhesion between the headband and the head or designing the headband geometry in such a way to constrain its movement. High-speed videography should be used in future tests to quantify the extent of headband sliding and develop methods to further remove its contribution from the head kinematics reconstruction. The effect of adjusting and repositioning the headband by an athlete during play on the estimated head kinematics and brain strains should also be studied in detail. This study presents the data-processing and results using the Storelli headband, and the results could be different for headbands with different material properties and designs.

The safety of the headband after instrumenting with sensors should also be investigated. 
When properly heading a soccer ball, the front of the head is typically impacted, so the sensors were placed toward the back of the head in this study to reduce the risk of directly impacting the sensors. However, 
there is a risk that other locations of the head may be impacted during a game.  Although the headband foam provides some cushioning, a direct impact to the sensors will cause a localized pressure point and should be avoided. Reducing the size of the sensors in future headband designs can minimize this pressure point and any potential risk of injury.

\section{Conclusions}
In this study, we developed an alternative wearable sensor system to instrumented mouthguards for activities where mouthguard use is either not feasible or has low participant compliance. 
A commercially available headband was instrumented with an array of IMU sensors and a new wavelet transform-based filtering approach was developed to improve the head rotational kinematics reconstruction from the sensors.
The headband performance was assessed in the laboratory under soccer headers, where repeated low magnitude head impacts have raised concern for potential brain injury. 
Our headband reconstruction of the angular velocity time histories showed `good' to `excellent' agreement (CORA scores) with the reference data for soccer ball impacts near the front of the head. The correlation coefficients for the peak rotational velocity and acceleration for these impacts were 0.80 and 0.63, respectively, and the NRMSE values were 0.20 and 0.28, respectively. The improved head kinematics data collection for impacts to the front of the head using instrumented headband can significantly enhance our ability to study head impact exposure in soccer for a large cohort.

\backmatter

\bmhead{Acknowledgments}
We are also grateful to Team Wendy for assisting with the laboratory experiments and to the PANTHER program for facilitating fruitful discussions and collaborations.

\section*{Declarations}

\begin{itemize}
\item \textbf{Funding:} The authors gratefully acknowledge the support from the University of Wisconsin-Madison Office of the Vice Chancellor for Research and Graduate Education (OVCRGE) and the Athletic Department. Funding for this award has been provided through Big 10 Athletic Media Revenue (136-AAI3375). The authors also acknowledge the U.S. Office of Naval Research funding under the PANTHER award N00014-21-1-2044 through Dr. Timothy Bentley.

\item \textbf{Competing interests:} The authors have no competing interests to declare that are relevant to the content of this article. 

\end{itemize}
\noindent

\bigskip

\begin{appendices}

\section{Wavelet Scalograms}\label{secA3}

The continuous wavelet transform (CWT) of a  signal $x(\tau)$ is defined as \cite{sadowsky1994continuous}
\begin{equation}
    w(\beta,\eta; x)=\frac{1}{\sqrt{\eta}}\int_{-\infty}^{\infty}x(\tau)\overline{\psi}\pr{\frac{\tau-\beta}{\eta}}{\rm d}\tau,
\label{eq:wbetaeta2}
\end{equation}
where $\beta\in\mathbb{R}$ is the shifting parameter and  $\eta\in\mathbb{R}$ is the scaling parameter. 
Here, we compute the CWT for a discrete set of $\beta$ and $\eta$ values: $\beta$ ranges over 
$[(p-1)\Delta\tau,\ldots,0,\Delta \tau,2\Delta\tau,\ldots,(n-p)\Delta\tau]$, 
where $\Delta \tau$ is the time difference between two consecutive signal readings; 
and $n$ is the number of times the signal was sampled; $p$ is the number of times the signal was sampled before the impact started; 
and $\eta$ ranges over the values of the 
function $\pr{{\rm oct},{\rm voc}}\mapsto \breve{\eta}\pr{{\rm oct},{\rm voc}}$,
\begin{equation}
\breve{\eta}\pr{{\rm oct},{\rm voc}}:=\alpha 2^{{\rm oct}-1}2^{{\rm voc}/40},
\end{equation} 
for ${\rm oct}\in \{1,2,\ldots,10\}$,  ${\rm voc}\in \{1,2,\ldots,40\}$, and $\alpha=1.92$.
The function $\psi(\cdot): \mathbb{R}\to \mathbb{C}$,  
\begin{equation}
\psi\pr{\tau}= \frac{e^{15\tau i } }{\pi^{\frac{1}{4}} e^{\frac{\tau^2}{2}}}, 
\label{eq:SelectedGaborWavelet}
\end{equation}
is a Gabor wavelet and $\overline{\psi}\pr{\frac{\tau-\beta}{\eta}}$ is the complex conjugate of $\psi\pr{\frac{\tau-\beta}{\eta}}$. 

\paragraph{Scalogram} 
The CWT can provide us the signal's scalogram, which can be described as the frequency spectrum of the signal at any given time. 
The quantity  $f(\eta)$ is 
    \begin{equation}
f(\eta):=\frac{f_c}{\eta \Delta\tau },
\label{eq:EtaFrequency}
\end{equation}
can be thought out as a frequency corresponding to the time scale $\eta$, and $f_c$ is the central frequency of $\psi(\cdot)$. For the $\psi(\cdot)$ given in Eq. \eqref{eq:SelectedGaborWavelet}, $f_c=2.39$. Let $\breve{w}(\cdot,\cdot;x): \mathbb{R} \times \mathbb{R}_{> 0} \to \mathbb{C}$, 
\begin{equation}
\breve{w}(\beta,f;x):=w(\beta,\frac{f_c}{\Delta\tau f};x).
\end{equation}
Here, $\breve{w}(\beta,f;x)$ is the wavelet co-efficient of the signal $x(\cdot)$ at the time instance $\beta$ and at the frequency scale $f$. 
The contour plot of the magnitude of $\breve{w}(\cdot,\cdot;x)$ is the signal $x(\cdot)$'s scalogram. The scalogram of a representative kinematic signals are shown in Figure \ref{fig4}c and d.

\section{Statistical Measures}\label{secA1}
The concordance correlation coefficient (CCC) for headband data against the reference measurement is given by 

\begin{equation}
    CCC = \frac{2\rho}{\mu + \frac{1}{\mu} + u^2}
\end{equation}\label{CCC}
where $\mu=\frac{S_x}{S_y}$ and $u= \frac{\bar{x} - \bar{y}}{\sqrt{S_x S_y}}$. The Pearson correlation coefficient $\rho$ is defined as
\begin{equation}
    \rho = \frac{ \sum_{i=1}^{n}(x_i-\bar{x})(y_i-\bar{y}) }{%
        \sqrt{\sum_{i=1}^{n}(x_i-\bar{x})^2}\sqrt{\sum_{i=1}^{n}(y_i-\bar{y})^2}}
\end{equation}
where $n$ is the number of data points, $x_i$ and $y_i$ represent the reference and headband data points, $\bar{x}$ and $\bar{y}$ represent their measurement means, and $S_x$ and $S_y$ represent their standard deviations.

The normalized root mean square error (NRMSE) of the headband data was calculated as
\begin{equation}
    NRMSE = \mathrm{\sqrt{\frac{\sum_{i=1} ^{n} (\frac{y_{i} - x_{i}}{x_{i}})^2}{n}}}
\end{equation}
where $x_i$ and $y_i$ represent the reference and headband data points, respectively.

\end{appendices}


\bibliography{sn-bibliography}


\begin{thebibliography}{53}
\ifx \bisbn   \undefined \def \bisbn  #1{ISBN #1}\fi
\ifx \binits  \undefined \def \binits#1{#1}\fi
\ifx \bauthor  \undefined \def \bauthor#1{#1}\fi
\ifx \batitle  \undefined \def \batitle#1{#1}\fi
\ifx \bjtitle  \undefined \def \bjtitle#1{#1}\fi
\ifx \bvolume  \undefined \def \bvolume#1{\textbf{#1}}\fi
\ifx \byear  \undefined \def \byear#1{#1}\fi
\ifx \bissue  \undefined \def \bissue#1{#1}\fi
\ifx \bfpage  \undefined \def \bfpage#1{#1}\fi
\ifx \blpage  \undefined \def \blpage #1{#1}\fi
\ifx \burl  \undefined \def \burl#1{\textsf{#1}}\fi
\ifx \doiurl  \undefined \def \doiurl#1{\url{https://doi.org/#1}}\fi
\ifx \betal  \undefined \def \betal{\textit{et al.}}\fi
\ifx \binstitute  \undefined \def \binstitute#1{#1}\fi
\ifx \binstitutionaled  \undefined \def \binstitutionaled#1{#1}\fi
\ifx \bctitle  \undefined \def \bctitle#1{#1}\fi
\ifx \beditor  \undefined \def \beditor#1{#1}\fi
\ifx \bpublisher  \undefined \def \bpublisher#1{#1}\fi
\ifx \bbtitle  \undefined \def \bbtitle#1{#1}\fi
\ifx \bedition  \undefined \def \bedition#1{#1}\fi
\ifx \bseriesno  \undefined \def \bseriesno#1{#1}\fi
\ifx \blocation  \undefined \def \blocation#1{#1}\fi
\ifx \bsertitle  \undefined \def \bsertitle#1{#1}\fi
\ifx \bsnm \undefined \def \bsnm#1{#1}\fi
\ifx \bsuffix \undefined \def \bsuffix#1{#1}\fi
\ifx \bparticle \undefined \def \bparticle#1{#1}\fi
\ifx \barticle \undefined \def \barticle#1{#1}\fi
\bibcommenthead
\ifx \bconfdate \undefined \def \bconfdate #1{#1}\fi
\ifx \botherref \undefined \def \botherref #1{#1}\fi
\ifx \url \undefined \def \url#1{\textsf{#1}}\fi
\ifx \bchapter \undefined \def \bchapter#1{#1}\fi
\ifx \bbook \undefined \def \bbook#1{#1}\fi
\ifx \bcomment \undefined \def \bcomment#1{#1}\fi
\ifx \oauthor \undefined \def \oauthor#1{#1}\fi
\ifx \citeauthoryear \undefined \def \citeauthoryear#1{#1}\fi
\ifx \endbibitem  \undefined \def \endbibitem {}\fi
\ifx \bconflocation  \undefined \def \bconflocation#1{#1}\fi
\ifx \arxivurl  \undefined \def \arxivurl#1{\textsf{#1}}\fi
\csname PreBibitemsHook\endcsname

\bibitem[\protect\citeauthoryear{Buck}{2011}]{buck2011mild}
\begin{barticle}
\bauthor{\bsnm{Buck}, \binits{P.W.}}:
\batitle{Mild traumatic brain injury: a silent epidemic in our practices}.
\bjtitle{Health \& social work}
\bvolume{36}(\bissue{4}),
\bfpage{299}--\blpage{302}
(\byear{2011})
\end{barticle}
\endbibitem

\bibitem[\protect\citeauthoryear{Stein et~al.}{2015}]{stein2015concussion}
\begin{barticle}
\bauthor{\bsnm{Stein}, \binits{T.D.}},
\bauthor{\bsnm{Alvarez}, \binits{V.E.}},
\bauthor{\bsnm{McKee}, \binits{A.C.}}:
\batitle{Concussion in chronic traumatic encephalopathy}.
\bjtitle{Current pain and headache reports}
\bvolume{19},
\bfpage{1}--\blpage{6}
(\byear{2015})
\end{barticle}
\endbibitem

\bibitem[\protect\citeauthoryear{Brett et~al.}{2022}]{brett2022traumatic}
\begin{barticle}
\bauthor{\bsnm{Brett}, \binits{B.L.}},
\bauthor{\bsnm{Gardner}, \binits{R.C.}},
\bauthor{\bsnm{Godbout}, \binits{J.}},
\bauthor{\bsnm{Dams-O’Connor}, \binits{K.}},
\bauthor{\bsnm{Keene}, \binits{C.D.}}:
\batitle{Traumatic brain injury and risk of neurodegenerative disorder}.
\bjtitle{Biological psychiatry}
\bvolume{91}(\bissue{5}),
\bfpage{498}--\blpage{507}
(\byear{2022})
\end{barticle}
\endbibitem

\bibitem[\protect\citeauthoryear{Zhan et~al.}{2021}]{Zhan2021}
\begin{botherref}
\oauthor{\bsnm{Zhan}, \binits{X.}},
\oauthor{\bsnm{Li}, \binits{Y.}},
\oauthor{\bsnm{Liu}, \binits{Y.}},
\oauthor{\bsnm{Domel}, \binits{A.G.}},
\oauthor{\bsnm{Alizadeh}, \binits{H.V.}},
\oauthor{\bsnm{Raymond}, \binits{S.J.}},
\oauthor{\bsnm{Ruan}, \binits{J.}},
\oauthor{\bsnm{Barbat}, \binits{S.}},
\oauthor{\bsnm{Tiernan}, \binits{S.}},
\oauthor{\bsnm{Gevaert}, \binits{O.}},
\oauthor{\bsnm{Zeineh}, \binits{M.M.}},
\oauthor{\bsnm{Grant}, \binits{G.A.}},
\oauthor{\bsnm{Camarillo}, \binits{D.B.}}:
The relationship between brain injury criteria and brain strain across different types of head impacts can be different.
Journal of the Royal Society Interface
\textbf{18}
(2021)
\doiurl{10.1098/rsif.2021.0260}
\end{botherref}
\endbibitem

\bibitem[\protect\citeauthoryear{Pellman et~al.}{2003}]{Pellman}
\begin{barticle}
\bauthor{\bsnm{Pellman}, \binits{E.J.}},
\bauthor{\bsnm{Viano}, \binits{D.C.}},
\bauthor{\bsnm{Tucker}, \binits{A.M.}},
\bauthor{\bsnm{Casson}, \binits{I.R.}},
\bauthor{\bsnm{Waeckerle}, \binits{J.F.}}:
\batitle{Concussion in professional football: Reconstruction of game impacts and injuries}.
\bjtitle{Neurosurgery}
\bvolume{53}(\bissue{4}),
\bfpage{799}--\blpage{814}
(\byear{2003})
\doiurl{10.1093/neurosurgery/53.3.799. PMID: 14519212.}
\end{barticle}
\endbibitem

\bibitem[\protect\citeauthoryear{Karton et~al.}{2016}]{Karton}
\begin{bchapter}
\bauthor{\bsnm{Karton}, \binits{C.}},
\bauthor{\bsnm{Oeur}, \binits{R.A.}},
\bauthor{\bsnm{Hoshizaki}, \binits{T.B.}}:
\bctitle{Measurement accuracy of head impact monitoring sensor in sport}.
In: \bbtitle{ISBS-Conference Proceedings Archive}
(\byear{2016}).
\burl{https://api.semanticscholar.org/CorpusID:55093276}
\end{bchapter}
\endbibitem

\bibitem[\protect\citeauthoryear{Buice et~al.}{2018}]{Buice2018}
\begin{botherref}
\oauthor{\bsnm{Buice}, \binits{J.M.}},
\oauthor{\bsnm{Esquivel}, \binits{A.O.}},
\oauthor{\bsnm{Andrecovich}, \binits{C.J.}}:
Laboratory validation of a wearable sensor for the measurement of head acceleration in men's and women's lacrosse.
Journal of Biomechanical Engineering
\textbf{140}
(2018)
\doiurl{10.1115/1.4040311}
\end{botherref}
\endbibitem

\bibitem[\protect\citeauthoryear{Tiernan et~al.}{2018}]{Tiernan2018}
\begin{barticle}
\bauthor{\bsnm{Tiernan}, \binits{S.}},
\bauthor{\bsnm{O’Sullivan}, \binits{D.}},
\bauthor{\bsnm{Byrne}, \binits{G.}}:
\batitle{Repeatability and reliability evaluation of a wireless head-band sensor}.
\bjtitle{The Asian Journal of Kinesiology}
\bvolume{20},
\bfpage{70}--\blpage{75}
(\byear{2018})
\doiurl{10.15758/ajk.2018.20.4.70}
\end{barticle}
\endbibitem

\bibitem[\protect\citeauthoryear{Cecchi et~al.}{2020}]{Cecchi2020}
\begin{barticle}
\bauthor{\bsnm{Cecchi}, \binits{N.J.}},
\bauthor{\bsnm{Monroe}, \binits{D.C.}},
\bauthor{\bsnm{Oros}, \binits{T.J.}},
\bauthor{\bsnm{Small}, \binits{S.L.}},
\bauthor{\bsnm{Hicks}, \binits{J.W.}}:
\batitle{Laboratory evaluation of a wearable head impact sensor for use in water polo and land sports}.
\bjtitle{Proceedings of the Institution of Mechanical Engineers, Part P: Journal of Sports Engineering and Technology}
\bvolume{234},
\bfpage{162}--\blpage{169}
(\byear{2020})
\doiurl{10.1177/1754337120901974}
\end{barticle}
\endbibitem

\bibitem[\protect\citeauthoryear{Patton et~al.}{2021}]{patton2021}
\begin{botherref}
\oauthor{\bsnm{Patton}, \binits{D.A.}},
\oauthor{\bsnm{Huber}, \binits{C.M.}},
\oauthor{\bsnm{Douglas}, \binits{E.C.}},
\oauthor{\bsnm{Seacrist}, \binits{T.}},
\oauthor{\bsnm{Arbogast}, \binits{K.B.}}:
Laboratory assessment of a head impact sensor for youth soccer ball heading impacts using an anthropomorphic test device.
Proceedings of the Institution of Mechanical Engineers, Part P: Journal of Sports Engineering and Technology,
17543371211063124
(2021)
\doiurl{10.1177/17543371211063124}
\end{botherref}
\endbibitem

\bibitem[\protect\citeauthoryear{Hanlon and Bir}{2010}]{hanlon2010validation}
\begin{barticle}
\bauthor{\bsnm{Hanlon}, \binits{E.}},
\bauthor{\bsnm{Bir}, \binits{C.}}:
\batitle{Validation of a wireless head acceleration measurement system for use in soccer play}.
\bjtitle{Journal of applied biomechanics}
\bvolume{26}(\bissue{4}),
\bfpage{424}--\blpage{431}
(\byear{2010})
\doiurl{10.1123/jab.26.4.424.}
\end{barticle}
\endbibitem

\bibitem[\protect\citeauthoryear{Sandmo et~al.}{2019}]{sandmo2019evaluation}
\begin{barticle}
\bauthor{\bsnm{Sandmo}, \binits{S.B.}},
\bauthor{\bsnm{McIntosh}, \binits{A.S.}},
\bauthor{\bsnm{Andersen}, \binits{T.E.}},
\bauthor{\bsnm{Koerte}, \binits{I.K.}},
\bauthor{\bsnm{Bahr}, \binits{R.}}:
\batitle{Evaluation of an in-ear sensor for quantifying head impacts in youth soccer}.
\bjtitle{The American journal of sports medicine}
\bvolume{47}(\bissue{4}),
\bfpage{974}--\blpage{981}
(\byear{2019})
\doiurl{10.1177/0363546519826953.}
\end{barticle}
\endbibitem

\bibitem[\protect\citeauthoryear{Schussler et~al.}{2017}]{schussler2017comparison}
\begin{barticle}
\bauthor{\bsnm{Schussler}, \binits{E.}},
\bauthor{\bsnm{Stark}, \binits{D.}},
\bauthor{\bsnm{Bolte}, \binits{J.H.}},
\bauthor{\bsnm{Kang}, \binits{Y.S.}},
\bauthor{\bsnm{Onate}, \binits{J.A.}}:
\batitle{Comparison of a head mounted impact measurement device to the hybrid iii anthropomorphic testing device in a controlled laboratory setting}.
\bjtitle{International journal of sports physical therapy}
\bvolume{12}(\bissue{4}),
\bfpage{592}
(\byear{2017})
\end{barticle}
\endbibitem

\bibitem[\protect\citeauthoryear{Kieffer et~al.}{2020}]{kieffer2020two}
\begin{barticle}
\bauthor{\bsnm{Kieffer}, \binits{E.E.}},
\bauthor{\bsnm{Begonia}, \binits{M.T.}},
\bauthor{\bsnm{Tyson}, \binits{A.M.}},
\bauthor{\bsnm{Rowson}, \binits{S.}}:
\batitle{A two-phased approach to quantifying head impact sensor accuracy: in-laboratory and on-field assessments}.
\bjtitle{Annals of biomedical engineering}
\bvolume{48},
\bfpage{2613}--\blpage{2625}
(\byear{2020})
\doiurl{10.1007/s10439-020-02647-1.}
\end{barticle}
\endbibitem

\bibitem[\protect\citeauthoryear{Stitt et~al.}{2021}]{stitt2021laboratory}
\begin{barticle}
\bauthor{\bsnm{Stitt}, \binits{D.}},
\bauthor{\bsnm{Draper}, \binits{N.}},
\bauthor{\bsnm{Alexander}, \binits{K.}},
\bauthor{\bsnm{Kabaliuk}, \binits{N.}}:
\batitle{Laboratory validation of instrumented mouthguard for use in sport}.
\bjtitle{Sensors}
\bvolume{21}(\bissue{18}),
\bfpage{6028}
(\byear{2021})
\doiurl{10.3390/s21186028.}
\end{barticle}
\endbibitem

\bibitem[\protect\citeauthoryear{Rich et~al.}{2019}]{rich2019development}
\begin{barticle}
\bauthor{\bsnm{Rich}, \binits{A.M.}},
\bauthor{\bsnm{Filben}, \binits{T.M.}},
\bauthor{\bsnm{Miller}, \binits{L.E.}},
\bauthor{\bsnm{Tomblin}, \binits{B.T.}},
\bauthor{\bsnm{Van~Gorkom}, \binits{A.R.}},
\bauthor{\bsnm{Hurst}, \binits{M.A.}},
\bauthor{\bsnm{Barnard}, \binits{R.T.}},
\bauthor{\bsnm{Kohn}, \binits{D.S.}},
\bauthor{\bsnm{Urban}, \binits{J.E.}},
\bauthor{\bsnm{Stitzel}, \binits{J.D.}}:
\batitle{Development, validation and pilot field deployment of a custom mouthpiece for head impact measurement}.
\bjtitle{Annals of biomedical engineering}
\bvolume{47},
\bfpage{2109}--\blpage{2121}
(\byear{2019})
\doiurl{10.1007/s10439-019-02313-1.}
\end{barticle}
\endbibitem

\bibitem[\protect\citeauthoryear{Miller et~al.}{2018}]{miller2018validation}
\begin{botherref}
\oauthor{\bsnm{Miller}, \binits{L.E.}},
\oauthor{\bsnm{Kuo}, \binits{C.}},
\oauthor{\bsnm{Wu}, \binits{L.C.}},
\oauthor{\bsnm{Urban}, \binits{J.E.}},
\oauthor{\bsnm{Camarillo}, \binits{D.B.}},
\oauthor{\bsnm{Stitzel}, \binits{J.D.}}:
Validation of a custom instrumented retainer form factor for measuring linear and angular head impact kinematics.
Journal of biomechanical engineering
\textbf{140}(5)
(2018)
\doiurl{10.1115/1.4039165. PMID: 29383374.}
\end{botherref}
\endbibitem

\bibitem[\protect\citeauthoryear{Kuo et~al.}{2016}]{kuo2016}
\begin{barticle}
\bauthor{\bsnm{Kuo}, \binits{C.}},
\bauthor{\bsnm{Wu}, \binits{L.C.}},
\bauthor{\bsnm{Hammoor}, \binits{B.T.}},
\bauthor{\bsnm{Luck}, \binits{J.F.}},
\bauthor{\bsnm{Cutcliffe}, \binits{H.C.}},
\bauthor{\bsnm{Lynall}, \binits{R.C.}},
\bauthor{\bsnm{Kait}, \binits{J.R.}},
\bauthor{\bsnm{Campbell}, \binits{K.R.}},
\bauthor{\bsnm{Mihalik}, \binits{J.P.}},
\bauthor{\bsnm{Bass}, \binits{C.R.}}, \betal:
\batitle{Effect of the mandible on mouthguard measurements of head kinematics}.
\bjtitle{Journal of biomechanics}
\bvolume{49}(\bissue{9}),
\bfpage{1845}--\blpage{1853}
(\byear{2016})
\doiurl{10.1016/j.jbiomech.2016.04.017.}
\end{barticle}
\endbibitem

\bibitem[\protect\citeauthoryear{McGuine et~al.}{2020}]{mcguine2020does}
\begin{barticle}
\bauthor{\bsnm{McGuine}, \binits{T.}},
\bauthor{\bsnm{Post}, \binits{E.}},
\bauthor{\bsnm{Pfaller}, \binits{A.Y.}},
\bauthor{\bsnm{Hetzel}, \binits{S.}},
\bauthor{\bsnm{Schwarz}, \binits{A.}},
\bauthor{\bsnm{Brooks}, \binits{M.A.}},
\bauthor{\bsnm{Kliethermes}, \binits{S.A.}}:
\batitle{Does soccer headgear reduce the incidence of sport-related concussion? a cluster, randomised controlled trial of adolescent athletes}.
\bjtitle{British journal of sports medicine}
\bvolume{54}(\bissue{7}),
\bfpage{408}--\blpage{413}
(\byear{2020})
\doiurl{10.1136/bjsports-2018-100238.}
\end{barticle}
\endbibitem

\bibitem[\protect\citeauthoryear{Hawn et~al.}{2002}]{hawn2002enforcement}
\begin{barticle}
\bauthor{\bsnm{Hawn}, \binits{K.L.}},
\bauthor{\bsnm{Visser}, \binits{M.F.}},
\bauthor{\bsnm{Sexton}, \binits{P.J.}}:
\batitle{Enforcement of mouthguard use and athlete compliance in national collegiate athletic association men's collegiate ice hockey competition}.
\bjtitle{Journal of athletic training}
\bvolume{37}(\bissue{2}),
\bfpage{204}
(\byear{2002})
\end{barticle}
\endbibitem

\bibitem[\protect\citeauthoryear{Boffano et~al.}{2012}]{boffano2012rugby}
\begin{barticle}
\bauthor{\bsnm{Boffano}, \binits{P.}},
\bauthor{\bsnm{Boffano}, \binits{M.}},
\bauthor{\bsnm{Gallesio}, \binits{C.}},
\bauthor{\bsnm{Roccia}, \binits{F.}},
\bauthor{\bsnm{Cignetti}, \binits{R.}},
\bauthor{\bsnm{Piana}, \binits{R.}}:
\batitle{Rugby athletes’ awareness and compliance in the use of mouthguards in the north west of italy}.
\bjtitle{Dental Traumatology}
\bvolume{28}(\bissue{3}),
\bfpage{210}--\blpage{213}
(\byear{2012})
\end{barticle}
\endbibitem

\bibitem[\protect\citeauthoryear{Matalon et~al.}{2008}]{matalon2008compliance}
\begin{barticle}
\bauthor{\bsnm{Matalon}, \binits{V.}},
\bauthor{\bsnm{Brin}, \binits{I.}},
\bauthor{\bsnm{Moskovitz}, \binits{M.}},
\bauthor{\bsnm{Ram}, \binits{D.}}:
\batitle{Compliance of children and youngsters in the use of mouthguards}.
\bjtitle{Dental traumatology}
\bvolume{24}(\bissue{4}),
\bfpage{462}--\blpage{467}
(\byear{2008})
\end{barticle}
\endbibitem

\bibitem[\protect\citeauthoryear{Liew et~al.}{2014}]{liew2014factors}
\begin{barticle}
\bauthor{\bsnm{Liew}, \binits{A.K.C.}},
\bauthor{\bsnm{Abdullah}, \binits{D.}},
\bauthor{\bsnm{Wan~Noorina}, \binits{W.A.}},
\bauthor{\bsnm{Khoo}, \binits{S.}}:
\batitle{Factors associated with mouthguard use and discontinuation among rugby players in m alaysia}.
\bjtitle{Dental Traumatology}
\bvolume{30}(\bissue{6}),
\bfpage{461}--\blpage{467}
(\byear{2014})
\end{barticle}
\endbibitem

\bibitem[\protect\citeauthoryear{Roberts}{2023}]{roberts2023sports}
\begin{barticle}
\bauthor{\bsnm{Roberts}, \binits{H.W.}}:
\batitle{Sports mouthguard overview: materials, fabrication techniques, existing standards, and future research needs}.
\bjtitle{Dental traumatology}
\bvolume{39}(\bissue{2}),
\bfpage{101}--\blpage{108}
(\byear{2023})
\end{barticle}
\endbibitem

\bibitem[\protect\citeauthoryear{Shore and O’connell}{2020}]{shoreinvestigation}
\begin{botherref}
\oauthor{\bsnm{Shore}, \binits{E.P.}},
\oauthor{\bsnm{O’connell}, \binits{A.}}:
An investigation into player compliance and level of protection afforded by mouthguards worn by children playing sport in ireland.
PhD thesis,
Trinity College, University of Dublin
(2020)
\end{botherref}
\endbibitem

\bibitem[\protect\citeauthoryear{Carey et~al.}{2021}]{carey2021video}
\begin{barticle}
\bauthor{\bsnm{Carey}, \binits{L.}},
\bauthor{\bsnm{Terry}, \binits{D.P.}},
\bauthor{\bsnm{McIntosh}, \binits{A.S.}},
\bauthor{\bsnm{Stanwell}, \binits{P.}},
\bauthor{\bsnm{Iverson}, \binits{G.L.}},
\bauthor{\bsnm{Gardner}, \binits{A.J.}}:
\batitle{Video analysis and verification of direct head impacts recorded by wearable sensors in junior rugby league players}.
\bjtitle{Sports Medicine-Open}
\bvolume{7},
\bfpage{1}--\blpage{13}
(\byear{2021})
\end{barticle}
\endbibitem

\bibitem[\protect\citeauthoryear{Tiernan et~al.}{2019}]{tiernan2019evaluation}
\begin{barticle}
\bauthor{\bsnm{Tiernan}, \binits{S.}},
\bauthor{\bsnm{Byrne}, \binits{G.}},
\bauthor{\bsnm{O’Sullivan}, \binits{D.M.}}:
\batitle{Evaluation of skin-mounted sensor for head impact measurement}.
\bjtitle{Proceedings of the Institution of Mechanical Engineers, Part H: Journal of Engineering in Medicine}
\bvolume{233}(\bissue{7}),
\bfpage{735}--\blpage{744}
(\byear{2019})
\end{barticle}
\endbibitem

\bibitem[\protect\citeauthoryear{Huber et~al.}{2021}]{huber2021laboratory}
\begin{barticle}
\bauthor{\bsnm{Huber}, \binits{C.M.}},
\bauthor{\bsnm{Patton}, \binits{D.A.}},
\bauthor{\bsnm{Wofford}, \binits{K.L.}},
\bauthor{\bsnm{Margulies}, \binits{S.S.}},
\bauthor{\bsnm{Cullen}, \binits{D.K.}},
\bauthor{\bsnm{Arbogast}, \binits{K.B.}}:
\batitle{Laboratory assessment of a headband-mounted sensor for measurement of head impact rotational kinematics}.
\bjtitle{Journal of Biomechanical Engineering}
\bvolume{143}(\bissue{2}),
\bfpage{024502}
(\byear{2021})
\doiurl{10.1115/1.4048574. PMID: 32975553.}
\end{barticle}
\endbibitem

\bibitem[\protect\citeauthoryear{Wu et~al.}{2016}]{wu2016bandwidth}
\begin{barticle}
\bauthor{\bsnm{Wu}, \binits{L.C.}},
\bauthor{\bsnm{Laksari}, \binits{K.}},
\bauthor{\bsnm{Kuo}, \binits{C.}},
\bauthor{\bsnm{Luck}, \binits{J.F.}},
\bauthor{\bsnm{Kleiven}, \binits{S.}},
\bauthor{\bsnm{Cameron}, \binits{R.}},
\bauthor{\bsnm{Camarillo}, \binits{D.B.}}:
\batitle{Bandwidth and sample rate requirements for wearable head impact sensors}.
\bjtitle{Journal of biomechanics}
\bvolume{49}(\bissue{13}),
\bfpage{2918}--\blpage{2924}
(\byear{2016})
\doiurl{10.1016/j.jbiomech.2016.07.004.}
\end{barticle}
\endbibitem

\bibitem[\protect\citeauthoryear{Kuo et~al.}{2018}]{kuo2018head}
\begin{barticle}
\bauthor{\bsnm{Kuo}, \binits{C.}},
\bauthor{\bsnm{Sganga}, \binits{J.}},
\bauthor{\bsnm{Fanton}, \binits{M.}},
\bauthor{\bsnm{Camarillo}, \binits{D.B.}}:
\batitle{Head impact kinematics estimation with network of inertial measurement units}.
\bjtitle{Journal of biomechanical engineering}
\bvolume{140}(\bissue{9}),
\bfpage{091006}
(\byear{2018})
\end{barticle}
\endbibitem

\bibitem[\protect\citeauthoryear{Wan et~al.}{2022}]{wan2022determining}
\begin{barticle}
\bauthor{\bsnm{Wan}, \binits{Y.}},
\bauthor{\bsnm{Fawzi}, \binits{A.L.}},
\bauthor{\bsnm{Kesari}, \binits{H.}}:
\batitle{Determining rigid body motion from accelerometer data through the square-root of a negative semi-definite tensor, with applications in mild traumatic brain injury}.
\bjtitle{Computer Methods in Applied Mechanics and Engineering}
\bvolume{390},
\bfpage{114271}
(\byear{2022})
\doiurl{10.1016/j.cma.2021.114271}
\end{barticle}
\endbibitem

\bibitem[\protect\citeauthoryear{Takhounts et~al.}{2013}]{takhounts2013development}
\begin{botherref}
\oauthor{\bsnm{Takhounts}, \binits{E.G.}},
\oauthor{\bsnm{Craig}, \binits{M.J.}},
\oauthor{\bsnm{Moorhouse}, \binits{K.}},
\oauthor{\bsnm{McFadden}, \binits{J.}},
\oauthor{\bsnm{Hasija}, \binits{V.}}:
Development of brain injury criteria ({B}r{IC}).
Technical report,
SAE Technical Paper
(2013)
\end{botherref}
\endbibitem

\bibitem[\protect\citeauthoryear{Zhang et~al.}{2006}]{zhang2006role}
\begin{barticle}
\bauthor{\bsnm{Zhang}, \binits{J.}},
\bauthor{\bsnm{Yoganandan}, \binits{N.}},
\bauthor{\bsnm{Pintar}, \binits{F.A.}},
\bauthor{\bsnm{Gennarelli}, \binits{T.A.}}:
\batitle{Role of translational and rotational accelerations on brain strain in lateral head impact}.
\bjtitle{Biomed Sci Instrum}
\bvolume{42},
\bfpage{501}--\blpage{506}
(\byear{2006})
\end{barticle}
\endbibitem

\bibitem[\protect\citeauthoryear{Ling et~al.}{2017}]{Ling2017}
\begin{barticle}
\bauthor{\bsnm{Ling}, \binits{H.}},
\bauthor{\bsnm{Morris}, \binits{H.R.}},
\bauthor{\bsnm{Neal}, \binits{J.W.}},
\bauthor{\bsnm{Lees}, \binits{A.J.}},
\bauthor{\bsnm{Hardy}, \binits{J.}},
\bauthor{\bsnm{Holton}, \binits{J.L.}},
\bauthor{\bsnm{Revesz}, \binits{T.}},
\bauthor{\bsnm{Williams}, \binits{D.D.R.}}:
\batitle{Mixed pathologies including chronic traumatic encephalopathy account for dementia in retired association football (soccer) players}.
\bjtitle{Acta Neuropathologica}
\bvolume{133},
\bfpage{337}--\blpage{352}
(\byear{2017})
\doiurl{10.1007/s00401-017-1680-3}
\end{barticle}
\endbibitem

\bibitem[\protect\citeauthoryear{Hales et~al.}{2014}]{Hales2307}
\begin{barticle}
\bauthor{\bsnm{Hales}, \binits{C.}},
\bauthor{\bsnm{Neill}, \binits{S.}},
\bauthor{\bsnm{Gearing}, \binits{M.}},
\bauthor{\bsnm{Cooper}, \binits{D.}},
\bauthor{\bsnm{Glass}, \binits{J.}},
\bauthor{\bsnm{Lah}, \binits{J.}}:
\batitle{Late-stage cte pathology in a retired soccer player with dementia}.
\bjtitle{Neurology}
\bvolume{83}(\bissue{24}),
\bfpage{2307}--\blpage{2309}
(\byear{2014})
\doiurl{10.1212/WNL.0000000000001081}
{\href{https://arxiv.org/abs/https://n.neurology.org/content/83/24/2307.full.pdf}{{https://n.neurology.org/content/83/24/2307.full.pdf}}}
\end{barticle}
\endbibitem

\bibitem[\protect\citeauthoryear{Ra\dj{}a et~al.}{2019}]{radja2019ball}
\begin{barticle}
\bauthor{\bsnm{Ra\dj{}a}, \binits{A.}},
\bauthor{\bsnm{Kuva{\v{c}}i{\'c}}, \binits{G.}},
\bauthor{\bsnm{De~Giorgio}, \binits{A.}},
\bauthor{\bsnm{Sellami}, \binits{M.}},
\bauthor{\bsnm{Ardig{\`o}}, \binits{L.P.}},
\bauthor{\bsnm{Bragazzi}, \binits{N.L.}},
\bauthor{\bsnm{Padulo}, \binits{J.}}:
\batitle{The ball kicking speed: A new, efficient performance indicator in youth soccer}.
\bjtitle{Plos one}
\bvolume{14}(\bissue{5}),
\bfpage{0217101}
(\byear{2019})
\end{barticle}
\endbibitem

\bibitem[\protect\citeauthoryear{Garc{\'\i}a-Pinillos et~al.}{2014}]{garcia2014effects}
\begin{barticle}
\bauthor{\bsnm{Garc{\'\i}a-Pinillos}, \binits{F.}},
\bauthor{\bsnm{Mart{\'\i}nez-Amat}, \binits{A.}},
\bauthor{\bsnm{Hita-Contreras}, \binits{F.}},
\bauthor{\bsnm{Mart{\'\i}nez-L{\'o}pez}, \binits{E.J.}},
\bauthor{\bsnm{Latorre-Rom{\'a}n}, \binits{P.A.}}:
\batitle{Effects of a contrast training program without external load on vertical jump, kicking speed, sprint, and agility of young soccer players}.
\bjtitle{The Journal of Strength \& Conditioning Research}
\bvolume{28}(\bissue{9}),
\bfpage{2452}--\blpage{2460}
(\byear{2014})
\end{barticle}
\endbibitem

\bibitem[\protect\citeauthoryear{Arbogast et~al.}{2022}]{arbogast2022consensus}
\begin{barticle}
\bauthor{\bsnm{Arbogast}, \binits{K.B.}},
\bauthor{\bsnm{Caccese}, \binits{J.B.}},
\bauthor{\bsnm{Buckley}, \binits{T.A.}},
\bauthor{\bsnm{McIntosh}, \binits{A.S.}},
\bauthor{\bsnm{Henderson}, \binits{K.}},
\bauthor{\bsnm{Stemper}, \binits{B.D.}},
\bauthor{\bsnm{Solomon}, \binits{G.}},
\bauthor{\bsnm{Broglio}, \binits{S.P.}},
\bauthor{\bsnm{Funk}, \binits{J.R.}},
\bauthor{\bsnm{Crandall}, \binits{J.R.}}:
\batitle{Consensus head acceleration measurement practices (champ): origins, methods, transparency and disclosure}.
\bjtitle{Annals of biomedical engineering}
\bvolume{50}(\bissue{11}),
\bfpage{1317}--\blpage{1345}
(\byear{2022})
\end{barticle}
\endbibitem

\bibitem[\protect\citeauthoryear{Hertz}{1899}]{hertz1899principles}
\begin{bbook}
\bauthor{\bsnm{Hertz}, \binits{H.}}:
\bbtitle{The Principles of Mechanics Presented in a New Form}.
\bpublisher{Macmillian and Company, Limited},
\blocation{New York}
(\byear{1899})
\end{bbook}
\endbibitem

\bibitem[\protect\citeauthoryear{Sadowsky}{1994}]{sadowsky1994continuous}
\begin{barticle}
\bauthor{\bsnm{Sadowsky}, \binits{J.}}:
\batitle{The continuous wavelet transform: A tool for signal investigation and understanding}.
\bjtitle{Johns Hopkins APL Technical Digest}
\bvolume{15},
\bfpage{306}--\blpage{306}
(\byear{1994})
\end{barticle}
\endbibitem

\bibitem[\protect\citeauthoryear{Lilly}{2017}]{lilly2017element}
\begin{barticle}
\bauthor{\bsnm{Lilly}, \binits{J.M.}}:
\batitle{Element analysis: A wavelet-based method for analysing time-localized events in noisy time series}.
\bjtitle{Proceedings of the Royal Society A: Mathematical, Physical and Engineering Sciences}
\bvolume{473}(\bissue{2200}),
\bfpage{20160776}
(\byear{2017})
\end{barticle}
\endbibitem

\bibitem[\protect\citeauthoryear{Grenke}{2002}]{grenke2002digital}
\begin{botherref}
\oauthor{\bsnm{Grenke}, \binits{B.D.}}:
Digital filtering for {J}211 requirements using a fast fourier transform based filter.
SAE Transactions,
359--401
(2002)
\end{botherref}
\endbibitem

\bibitem[\protect\citeauthoryear{Wolfram}{2010}]{WolframResearch}
\begin{botherref}
\oauthor{\bsnm{Wolfram}}:
Continuous wavelet transform.
Wolfram Language function
(2010)
\end{botherref}
\endbibitem

\bibitem[\protect\citeauthoryear{MATLAB}{2020}]{MATLAB:2020}
\begin{bbook}
\bauthor{\bsnm{MATLAB}}:
\bbtitle{Version 9.8.0.1417392 (R2020a)}.
\bpublisher{The MathWorks Inc.},
\blocation{Natick, Massachusetts}
(\byear{2020})
\end{bbook}
\endbibitem

\bibitem[\protect\citeauthoryear{Lawrence and Lin}{1989}]{lawrence1989concordance}
\begin{botherref}
\oauthor{\bsnm{Lawrence}, \binits{I.}},
\oauthor{\bsnm{Lin}, \binits{K.}}:
A concordance correlation coefficient to evaluate reproducibility.
Biometrics,
255--268
(1989)
\end{botherref}
\endbibitem

\bibitem[\protect\citeauthoryear{Giordano and Kleiven}{2016}]{giordano2016development}
\begin{botherref}
\oauthor{\bsnm{Giordano}, \binits{C.}},
\oauthor{\bsnm{Kleiven}, \binits{S.}}:
Development of an unbiased validation protocol to assess the biofidelity of finite element head models used in prediction of traumatic brain injury.
Technical report,
SAE Technical Paper
(2016)
\end{botherref}
\endbibitem

\bibitem[\protect\citeauthoryear{Gehre et~al.}{2009}]{gehre2009objective}
\begin{bchapter}
\bauthor{\bsnm{Gehre}, \binits{C.}},
\bauthor{\bsnm{Gades}, \binits{H.}},
\bauthor{\bsnm{Wernicke}, \binits{P.}}:
\bctitle{Objective rating of signals using test and simulation responses}.
In: \bbtitle{21st International Technical Conference on the Enhanced Safety of Vehicles Conference (ESV)},
pp. \bfpage{09}--\blpage{0407}
(\byear{2009})
\end{bchapter}
\endbibitem

\bibitem[\protect\citeauthoryear{Brooks et~al.}{2021}]{Brooks2021}
\begin{barticle}
\bauthor{\bsnm{Brooks}, \binits{J.S.}},
\bauthor{\bsnm{Allison}, \binits{W.}},
\bauthor{\bsnm{Harriss}, \binits{A.}},
\bauthor{\bsnm{Bian}, \binits{K.}},
\bauthor{\bsnm{Mao}, \binits{H.}},
\bauthor{\bsnm{Dickey}, \binits{J.P.}}:
\batitle{Purposeful heading performed by female youth soccer players leads to strain development in deep brain structures}.
\bjtitle{Neurotrauma Reports}
\bvolume{2},
\bfpage{354}--\blpage{362}
(\byear{2021})
\doiurl{10.1089/neur.2021.0014}
\end{barticle}
\endbibitem

\bibitem[\protect\citeauthoryear{Harriss et~al.}{2019}]{Harriss2019}
\begin{barticle}
\bauthor{\bsnm{Harriss}, \binits{A.}},
\bauthor{\bsnm{Johnson}, \binits{A.M.}},
\bauthor{\bsnm{Walton}, \binits{D.M.}},
\bauthor{\bsnm{Dickey}, \binits{J.P.}}:
\batitle{Head impact magnitudes that occur from purposeful soccer heading depend on the game scenario and head impact location}.
\bjtitle{Musculoskeletal Science and Practice}
\bvolume{40},
\bfpage{53}--\blpage{57}
(\byear{2019})
\doiurl{10.1016/j.msksp.2019.01.009}
\end{barticle}
\endbibitem

\bibitem[\protect\citeauthoryear{Chu et~al.}{2006}]{chu2006novel}
\begin{botherref}
\oauthor{\bsnm{Chu}, \binits{J.}},
\oauthor{\bsnm{Beckwith}, \binits{J.}},
\oauthor{\bsnm{Crisco}, \binits{J.}},
\oauthor{\bsnm{Greenwald}, \binits{R.}}:
A novel algorithm to measure linear and rotational head acceleration using single-axis accelerometers.
Journal of biomechanics
(39),
534
(2006)
\end{botherref}
\endbibitem

\bibitem[\protect\citeauthoryear{Rowson et~al.}{2022}]{rowson2022consensus}
\begin{barticle}
\bauthor{\bsnm{Rowson}, \binits{S.}},
\bauthor{\bsnm{Mihalik}, \binits{J.}},
\bauthor{\bsnm{Urban}, \binits{J.}},
\bauthor{\bsnm{Schmidt}, \binits{J.}},
\bauthor{\bsnm{Marshall}, \binits{S.}},
\bauthor{\bsnm{Harezlak}, \binits{J.}},
\bauthor{\bsnm{Stemper}, \binits{B.D.}},
\bauthor{\bsnm{McCrea}, \binits{M.}},
\bauthor{\bsnm{Funk}, \binits{J.}}:
\batitle{Consensus head acceleration measurement practices (champ): study design and statistical analysis}.
\bjtitle{Annals of Biomedical Engineering}
\bvolume{50}(\bissue{11}),
\bfpage{1346}--\blpage{1355}
(\byear{2022})
\end{barticle}
\endbibitem

\bibitem[\protect\citeauthoryear{Kuo et~al.}{2022}]{kuo2022field}
\begin{barticle}
\bauthor{\bsnm{Kuo}, \binits{C.}},
\bauthor{\bsnm{Patton}, \binits{D.}},
\bauthor{\bsnm{Rooks}, \binits{T.}},
\bauthor{\bsnm{Tierney}, \binits{G.}},
\bauthor{\bsnm{McIntosh}, \binits{A.}},
\bauthor{\bsnm{Lynall}, \binits{R.}},
\bauthor{\bsnm{Esquivel}, \binits{A.}},
\bauthor{\bsnm{Daniel}, \binits{R.}},
\bauthor{\bsnm{Kaminski}, \binits{T.}},
\bauthor{\bsnm{Mihalik}, \binits{J.}}, \betal:
\batitle{On-field deployment and validation for wearable devices}.
\bjtitle{Annals of biomedical engineering}
\bvolume{50}(\bissue{11}),
\bfpage{1372}--\blpage{1388}
(\byear{2022})
\doiurl{10.1007/s10439-022-03001-3}
\end{barticle}
\endbibitem

\bibitem[\protect\citeauthoryear{Rahaman et~al.}{2020}]{rahaman2020accelerometer}
\begin{barticle}
\bauthor{\bsnm{Rahaman}, \binits{M.M.}},
\bauthor{\bsnm{Fang}, \binits{W.}},
\bauthor{\bsnm{Fawzi}, \binits{A.L.}},
\bauthor{\bsnm{Wan}, \binits{Y.}},
\bauthor{\bsnm{Kesari}, \binits{H.}}:
\batitle{An accelerometer-only algorithm for determining the acceleration field of a rigid body, with application in studying the mechanics of mild traumatic brain injury}.
\bjtitle{Journal of the Mechanics and Physics of Solids}
\bvolume{143},
\bfpage{104014}
(\byear{2020})
\doiurl{10.1016/j.jmps.2020.104014}
\end{barticle}
\endbibitem

\end{thebibliography}

\end{document}